\documentclass[aps,prd,twocolumn,amsmath,superscriptaddress,amssymb,showpacs,floatfix,nofootinbib,longbibliography]{revtex4-1}
\usepackage{graphicx}
\usepackage{dcolumn}
\usepackage{xcolor}
\usepackage{bm}
\usepackage{natbib}
\usepackage{calligra}
\usepackage[T1]{fontenc}
\usepackage{egothic}
\usepackage[T1]{fontenc}
\newfont{\rsfsten}{rsfs10 scaled 1200}
\newfont{\rsfsseven}{rsfs10 scaled 1200}
\newfont{\rsfsfive}{rsfs10 scaled 1200}
\usepackage{epsfig}
\usepackage{units}
\usepackage[utf8]{inputenc}
\usepackage{hyperref}

\newcommand{\be}{\begin{equation}}
\newcommand{\ee}{\end{equation}}
\newcommand{\bea}{\begin{eqnarray}}
\newcommand{\eea}{\end{eqnarray}}

\def\lsim{\mathrel{\raise.3ex\hbox{$<$\kern-.75em\lower1ex\hbox{$\sim$}}}}
\def\gsim{\mathrel{\raise.3ex\hbox{$>$\kern-.75em\lower1ex\hbox{$\sim$}}}}

\begin{document}

\title{Limits on primordial black holes from the extragalactic gamma-ray background: Current status and future projections}

\author{Ilias Cholis}
\email{cholis@oakland.edu, ORCID: orcid.org/0000-0002-3805-6478}
\affiliation{Department of Physics, Oakland University, Rochester, Michigan, 48309, USA}
\author{Iason Krommydas}
\email{ik23@rice.edu, ORCID: orcid.org/0000-0001-7849-8863}
\affiliation{Department of Physics and Astronomy, Rice University, Houston, Texas, 77005, USA}
\author{John Carlini}
\email{jcarlini@oakland.edu, ORCID: orcid.org/0009-0002-2918-0300}
\affiliation{Department of Physics, Oakland University, Rochester, Michigan, 48309, USA}
\date{\today}

\begin{abstract}

Primordial black holes (PBHs), possibly formed from
the collapse of early universe perturbations, will evaporate via Hawking radiation with a lifetime comparable to the age of the universe, 
if their mass is $O(10^{14})$ g. Such black holes can contribute to the observed gamma-ray fluxes in the MeV and GeV range. 
Using the observed extragalactic gamma-ray background (EGRB) from the \textit{Fermi} Large Area Telescope, the \textit{EGRET}, and the \textit{COMPTEL} telescopes that cover gamma-ray 
energies from 0.5 MeV to 1 TeV, we evaluate limits on the abundance of PBHs with masses of $10^{14}$ to $10^{17}$ g. We study both monochromatic and extended mass distributions of PBHs. 
To model the EGRB spectrum, we calculate the contribution from extragalactic sources including blazars, star-forming galaxies and radio galaxies and also account 
for ultra-high-energy cosmic rays that produce gamma rays when interacting with the infrared background. Our EGRB modeling uses information from the \textit{Fermi} gamma-ray point sources catalog, from  observations at X-rays, the visible spectrum, the infrared and radio waves, and also accounts for modeling uncertainties and variations on the properties within each class of these sources. 
Moreover, we use recent work on the modeling of the PBHs' gamma-ray emission, that
includes the direct Hawking radiation, gamma rays produced in the hadronization and decay of unstable particles, final state radiation and gamma rays from pair annihilations in the interstellar 
medium. 
As the contribution of final state radiation and the annihilation of positrons enhances the low-energy part of the produced gamma-ray spectra from PBHs, we find that the EGRB observations can set the tightest limits on their abundance among all indirect dark matter probes, within the mass range of interest.  
PBHs with masses $\simeq 10^{14}$ g cannot contribute more than $O(10^{-10})$ of the observed dark matter. 
Finally, we discuss the impact of future observations from AMEGO-X and e-ASTROGAM, finding that these future MeV-scale telescopes can improve the current limits on PBHs of masses greater than $5 \times 10^{14}$ g by up to a factor of one hundred. 

\end{abstract}

\maketitle

\section{Introduction}
\label{sec:introduction}

Primordial black holes (PBHs) could have formed from the collapse of large primordial perturbations in the early universe through a
variety of mechanisms \cite{1967SvA....10..602Z, 1971MNRAS.152...75H, 1975Natur.253..251C, Carr:2016drx}. If PBHs do exist, they contribute to the observed dark matter abundance providing an answer to one of the most pressing questions in cosmology; the nature of dark matter.  

No PBHs have been detected so far, while there is a wide range of possible mechanisms for their formation. Thus, their mass 
range remains an open question. Limits on the PBH abundance, have been derived from a variety of observations \cite{Carr:2020gox}. 
For PBHs with masses up to $10^{17}$ g, limits have been derived from measurements of the abundance of nuclei 
produced during the Big Bang Nucleosynthesis \cite{1977SvAL....3..110Z, 1978SvAL....4..185V, 1980MNRAS.193..593L, Carr:2009jm}, 
from studying spectral distortions on the Cosmic Microwave Background \cite{Acharya:2020jbv, Chluba:2020oip}, from radio observations 
toward the galactic center \cite{Chan:2020zry}, from the gamma-ray background and galactic gamma-ray studies 
\cite{Carr:2009jm, Carr:2016hva, DeRocco:2019fjq, Laha:2019ssq, Arbey:2019vqx, Ballesteros:2019exr, Laha:2020ivk, Coogan:2020tuf, Ray:2021mxu}, from the heating of the 
interstellar medium of dwarf galaxies \cite{Kim:2020ngi, Lu:2020bmd} and from measuring the low-energy cosmic-ray 
electron and positron fluxes with \textit{Voyager 1} \cite{Boudaud:2018hqb}. 

For PBHs more massive than $10^{17}$ g, limits on their abundance can be derived 
from the impact close flybys of PBHs would have on the orbital trajectories of Solar System objects \cite{Tran:2023jci}, 
by PBHs causing femtolensing on the spectra of distant gamma-ray bursts \cite{2012PhRvD..86d3001B, Katz:2018zrn, Fedderke:2024wpy}, 
from microlensing of stars in Andromeda changing their observed flux \cite{Niikura:2017zjd, Sugiyama:2019dgt, Montero-Camacho:2019jte, Smyth:2019whb, Croon:2020ouk}, 
from searching for the same signature in observations in the \textit{Kepler} field of view \cite{Griest:2013aaa}, 
from OGLE microlensing searches of stars in the Galactic Bulge \cite{Niikura:2019kqi}, 
from EROS observations toward stars from the Large and Small Magellanic Clouds \cite{EROS-2:2006ryy} (see also \cite{Verma:2022pym}).
Also limits can be derived from the microlensing of X-ray pulsars \cite{Bai:2018bej}, and supernovae type Ia 
\cite{Zumalacarregui:2017qqd}, from analyzing the magnification of lensed objects \cite{Oguri:2017ock}, 
from the mergers of asteroid mass and planetary mass PBHs giving a signal/excess contribution to the continuous 
gravitational waves observed by LIGO \cite{Miller:2021knj}, from Lyman-$\alpha$ forest observations \cite{Murgia:2019duy}, 
from future stochastic gravitational-wave background 
observations \cite{Kuhnel:2018mlr, Sugiyama:2020roc}, from their lensing effect on fast radio bursts' received pulse \cite{Munoz:2016tmg, Laha:2018zav, Liao:2020wae, 2024ApJ...962...11Z}. 
PBHs can also affect dynamically either systems of binary stars leading to constraints on their abundance in the 
Milky Way \cite{2014ApJ...790..159M}, and affect the evolution of Milky Way globular clusters \cite{Brandt:2016aco,  Koushiappas:2017chw}.
The observations of black hole mergers by LIGO-Virgo and KAGRA \cite{LIGOScientific:2016aoc, KAGRA:2021duu}, has also led to the evaluation of 
the merger rate of PBH binaries \cite{Bird:2016dcv, Sasaki:2016jop, Ali-Haimoud:2017rtz, Raidal:2024bmm, Aljaf:2025dta}, and subsequently to limits on PBHs in the $O(10) \, M_{\odot}$ mass range \cite{Sasaki:2016jop, Ali-Haimoud:2017rtz, Kavanagh:2018ggo, Bouhaddouti:2025ltb, Bouhaddouti:2026jgc}. 
Finally, PBHs more massive than $O(100) \, M_{\odot}$ can cause detectable CMB temperature and polarization fluctuations, thus the \textit{Planck} measurements of the relevant power spectra \cite{Planck:2018vyg, Planck:2019nip}, can be used to set limits on their abundance \cite{Ricotti:2007au, Chen:2016pud, Ali-Haimoud:2016mbv, Horowitz:2016lib, Poulin:2017bwe, Serpico:2020ehh}.

In this paper we focus on the limits that can be derived from Hawking radiation \cite{1974Natur.248...30H, Hawking:1975vcx}, 
using the existing observations of the extragalactic gamma-ray background (EGRB) from the Solar Maximum Mission (\textit{SMM}) \cite{SMM1997}, 
from the \textit{Compton} Telescope (\textit{COMPTEL}) instrument onboard the \textit{Compton} Gamma Ray Observatory (\textit{CGRO}) 
\cite{10.1063/1.1307028}, from the Energetic Gamma-Ray Experiment Telescope (\textit{EGRET}) \cite{1998ApJ...494..523S}, 
and from the \textit{Fermi} Large Area Telescope (\textit{Fermi}-LAT) \cite{Fermi-LAT:2014ryh}. 
The combined emission from PBHs in dark matter halos of all sizes and integrated over redshift, can give an emission that would be nearly isotropic 
once focusing on latitudes away from Milky Way's disk. We also discuss the future limits that can 
be derived from future experiments in the MeV gamma-ray energy range, in particular from the All-sky Medium Energy Gamma-ray 
Observatory (AMEGO-X) \cite{AMEGO:2019gny} and the e-ASTROGAM telescope \cite{e-ASTROGAM:2017pxr}. We make use of the
new tools to calculate with great accuracy the full gamma-ray emission spectrum from the evaporation of black holes \cite{Arbey:2021mbl, 
Carlini:2025bki}. In particular we make use of the \texttt{GammaPBHPlotter} public code \cite{Carlini:2025bki}, that accounts for
 the direct Hawking emission component, for the secondary emission from the decay and hadronization of unstable particles relying 
 also on \cite{Arbey:2021mbl}, but also includes the final state radiation of relativistic particles, and the gamma-ray emission from the 
 in-flight annihilation of positrons radiated from the black holes with the interstellar medium. 
 
The EGRB is the combined emission of a vast number of resolved and unresolved sources; 
in particular active galactic nuclei (AGN)~\cite{Stecker:1993ni,1995MNRAS.277.1477P,  Salamon:1994ku,Stecker:1996ma,Mukherjee:1999it,Narumoto:2006qg,Giommi:2005bp,Dermer:2006pd,Pavlidou:2007dv, Inoue:2008pk, Ajello:2013lka, 
DiMauro:2013zfa, Qu:2019zln, Zeng:2021wln, Korsmeier:2022cwp},
radio galaxies~\cite{2011ApJ...733...66I, 
Stecker:2010di, Inoue:2011bm, DiMauro:2013xta, Hooper:2016gjy, Fukazawa:2022gwm} and star forming galaxies~\cite{Pavlidou:2002va,Thompson:2006qd,Fields:2010bw,Makiya:2010zt, Stecker:2010di, Chakraborty:2012sh, Lacki:2012si, Tamborra:2014xia, Fornasa:2015qua, Linden:2016fdd, Blanco:2021icw}, but also includes gamma rays from cascades produced 
as a result of  ultra-high-energy cosmic rays (UHECR) interacting with low-energy photons in the intergalactic medium
\cite{1973Natur.241..109S, 1973Ap&SS..20...47S, Cholis:2013ena, Globus:2017ehu, Aloisio:2017iyh, AlvesBatista:2019rhs}. 
The modeling of all these populations of sources has gradually improved with more extensive observations from gamma-ray instruments. 
Most importantly, \textit{Fermi}-LAT~\cite{Fermi-LAT:2010pat}, combined with observations in radio and the infrared has provided us with the possibility to do detailed population models, to account for the 
spectral properties, redshift distribution and luminosity functions of each of those classes of sources~\cite{Abazajian:2010pc, 2011ApJ...743..171A, Fermi-LAT:2015otn, Manconi:2019ynl, Malyshev:2011zi, Fermi-LAT:2015otn, Zechlin:2015wdz, Lisanti:2016jub, Fermi-LAT:2010tsy, Ajello:2011zi, Cuoco:2012yf, Harding:2012gk, Cholis:2013ena, Fornasa:2015qua, DiMauro:2015tfa, Stecker:2010di, Fermi-LAT:2012nqz, Inoue:2011bm, Cavadini:2011ig, Linden:2016fdd, Hooper:2016gjy, Blanco:2021icw, Ahlers:2011sd, Gelmini:2011kg, AlvesBatista:2019rhs}.

In section~\ref{sec:data}, we describe the EGRB spectral data used in this analysis as well as the data we expect to have from the AMEGO-X and the e-ASTROGAM future telescopes. 
Following, in section~\ref{sec:astro_backgrounds},
we briefly discuss the population modeling of known astrophysical components to the EGRB. We build upon previous work from Ref.~\cite{Cholis:2024hmd}, that focused 
on the isotropic gamma-ray background spectrum from 0.1 to 100 GeV. 
While the EGRB and the isotropic gamma-ray background are separate spectra, their modeling is closely related as we describe. 
In section~\ref{subsec:DM}, we describe how we 
model the contribution from monochromatic PBHs in the range of $1 \times 10^{14} - 3 \times 10^{17}$ g, 
to the EGRB, and derive limits on their abundance based on the current measurements from 1 MeV to 820 GeV. We also discuss the case of an extended mass distribution for PBHs following Ref.~\cite{Biagetti:2021eep}. 
We show in section~\ref{sec:PBHlimits}, that the current EGRB spectrum, can set limits on the abundance of monochromatic PBHs -by total mass contribution to the cosmological dark matter abundance- as massive as $7 \times 10^{16}$ g. These limits become very strong with decreasing PBH mass and can be as stringent as one part in $10^{10}$ for $1 \times 10^{14}$ g PBHs. Then in section~\ref{sec:Projections}, we discuss how the future gamma-ray observations in the MeV-range by AMEGO-X and e-ASTROGAM can 
affect our modeling of the EGRB spectrum and subsequently the limits on the PBH abundance. These future telescopes can improve the limits by up to a factor of one hundred. We give our final conclusions in section~\ref{sec:conclusions}. 

\section{The Extragalactic Gamma-Ray Background Spectrum}
\label{sec:data}

We use the \textit{Fermi}-LAT EGRB spectrum of Ref.~\cite{Fermi-LAT:2014ryh}, using the entire spectral data from 100 MeV to 820 GeV.
The EGRB and the isotropic gamma-ray background spectra used observations from medium and high galactic latitudes, with the isotropic spectrum excluding the emission from the resolved point sources and also removing the 
galactic diffuse emission. Instead, the EGRB includes the emission from the resolved point sources identified as extragalactic. 
These contributions are model dependent especially due to foreground model uncertainties. From the three alternative 
models for the foreground emission of Ref.~\cite{Fermi-LAT:2014ryh},  we used  model ``A'' that has the lowest flux at $\gtrsim 1$ GeV.  
We also used the EGRB spectra from \textit{EGRET}~\cite{1998ApJ...494..523S}, \textit{COMPTEL} \cite{10.1063/1.1307028} and compare to the EGRB data from \textit{SMM} \cite{SMM1997}, 
which we show in Fig.~\ref{fig:BackgroundModel}.
We chose to use the \textit{Fermi} EGRB spectrum as that measurement can be directly compared to the measurements from earlier telescopes. 
The \textit{Fermi}-LAT isotropic gamma-ray background spectrum has a very large fraction of the
extragalactic sources flux subtracted, as many of these sources have been identified as point sources, whereas the \textit{EGRET},\textit{COMPTEL} and \textit{SMM} observations do not. 
Using the EGRB data in our fitting procedure we can combine the \textit{Fermi}-LAT spectral data with the \textit{EGRET}, and \textit{COMPTEL} observations.

Furthermore, we make projections on the future observations from AMEGO-X and e-ASTROGAM. For these instruments we first take the quoted 
sensitivities for continuous emission from \cite{Fleischhack:2021mhc} for AMEGO-X and \cite{e-ASTROGAM:2017pxr} for e-ASTROGAM; 
which also compare their sensitivities to that of \textit{Fermi}-LAT. We use the averaged sky exposure from \textit{Fermi}-LAT at 10 years of 
observations for the \texttt{CLEAN} class of \textit{Fermi} events, to more directly compare to the assumptions of Refs.~\cite{Fleischhack:2021mhc, 
e-ASTROGAM:2017pxr}, and calculated projected spectra at three years of observations from  AMEGO-X and one-year observations from 
e-ASTROGAM. However, we note that in any realistic future evaluation of the EGRB from AMEGO-X or e-ASTROGAM, the main source of uncertainty in this spectrum is not going to be purely the systematic errors associated with the observed counts of photons. Instead, the systematics of the analysis are going to be the dominant cause of uncertainty on the derived EGRB spectrum. Those systematics, as has been the case with the EGRB spectrum from the \textit{Fermi} Collaboration, are related to removing the galactic diffuse emission and the emission from galactic point and extended sources from the total observed gamma-ray spectrum. 

We take that the future EGRB spectrum will be evaluated with a systematic fractional uncertainty of $7\%$ from either experiment. This is a somewhat conservative assumption on the potential of future analysis, as the \textit{Fermi} EGRB spectrum is evaluated with a fractional uncertainty of $\simeq 8\%$ between 1 and 10 GeV. Changing this assumption by a reasonable degree does not change our projected limits by more than a factor of 2. 

\section{Modeling the astrophysical background sources to the EGRB}
\label{sec:astro_backgrounds}

The EGRB is the result of the combined emission from extragalactic sources but may have a small contribution from galactic gamma-ray sources 
not bright enough to be identified as individual sources in relevant catalogues \cite{Fermi-LAT:2009ihh, Fermi-LAT:2011iqa, Fermi-LAT:2015bhf, 
Fermi-LAT:2019yla, Ballet:2023qzs}. The \textit{Fermi} EGRB is evaluated at galactic latitudes of $|b|> 20^{\circ}$,
after subtracting the galactic diffuse foreground emission \cite{Fermi-LAT:2014ryh}. Thus, most of its intensity 
is extragalactic. These sources include BL Lacertae (BL Lac) objects, flat-spectrum radio quasars (FSRQs), 
star-forming galaxies, radio galaxies and ultra-high-energy cosmic rays (UHECRs) interacting with the intergalactic medium 
\cite{Bringmann:2013ruh, Cholis:2013ena, DiMauro:2015tfa, Ajello:2015mfa, Fornasa:2015qua, Cholis:2024hmd}. At 
\textit{Fermi} gamma-ray energies a minor contribution can come from galactic millisecond pulsars (MSPs) (see e.g. 
Ref.~\cite{Cholis:2024hmd} for the latest analysis). 
We model separately each component from these classes of sources with their relevant astrophysical uncertainties 
and fit their combination to the EGRB spectrum.  

\vspace{-0.3cm}
\subsection{BL Lacertae objects and Flat-Spectrum Radio Quasars}
\label{subsec:BLLac-FSRQ}

BL Lacs and FSRQs are the two basic types of AGNs. Combined they are the most numerous 
classes of identified extragalactic gamma-ray sources ~\cite{Fermi-LAT:2019yla, Ballet:2023qzs}.  
Almost all AGNs emitting in gamma rays 
have been detected as sources by the \textit{Fermi}-LAT collaboration, but not by the earlier telescopes \textit{SMM}, \textit{COMPTEL} and \textit{EGRET}
(see Refs.~\cite{Ballet:2023qzs, 4FGL-DR4} for the latest update). 
By using the EGRB data from \textit{Fermi}-LAT, we allow for a direct connection between the observations from all telescopes and comparison to the EGRB models.

For the case of the EGRB models to be fitted to the \textit{Fermi} data, we include both objects that lie below the \textit{Fermi} detection threshold and the contribution of the detected objects.
The detected objects from these classes have 
measured gamma-ray spectra and have been discovered at different redshifts and wavelengths. We rely on 
our earlier work from Ref.~\cite{Cholis:2024hmd}, that used redshift and luminosity distributions for the BL Lacs and FSRQs~\cite{Ajello:2013lka, 
DiMauro:2013zfa, Qu:2019zln, Zeng:2021wln, Korsmeier:2022cwp}. 
The gamma-ray luminosity function for these objects is~\cite{Korsmeier:2022cwp}, 
\begin{equation}
\Phi (L_{\gamma}, \Gamma, z) = \Phi (L_{\gamma}, \Gamma, 0) \times e(L_{\gamma}, z),
\label{eq:AGN_GLF}
\end{equation}
where $\Gamma$ is the photon spectral index, and $L_{\gamma}$ is the object's bolometric gamma-ray luminosity between 0.1 and 100 GeV energies.
For $\Phi (L_{\gamma}, \Gamma, 0)$, we use the parametrization introduced in~\cite{Korsmeier:2022cwp},  with the updates 
on parameter values from fitting the \textit{Fermi} EGRB spectrum from Ref.~\cite{Cholis:2024hmd}, where a full description of the 
BL Lac and FSRQ luminosity functions modeling and underlying values for their parameterizations is provided. 

High energy gamma rays from extragalactic sources can interact with the intergalactic background light (especially the ultraviolet photons) 
leading to electron-positron pair production. This leads to an attenuation of their observed spectra at energies greater than $\simeq 50$ GeV. 
This attenuation can be expressed as an exponential suppression of the flux $exp \left[ - \tau(z, E_{\gamma}) \right] $, where 
$\tau(z, E_{\gamma})$ is the optical depth. We use the ``fiducial'' model from Ref.~\cite{2012MNRAS.422.3189G}. Since our fits of
the combination of  extragalactic populations of sources are dominated by the observations at energies lower than 50 GeV, the exact 
assumptions on the optical depth are not important. 

The total intensity of gamma rays from a class of extragalactic gamma-ray sources is,
\begin{eqnarray}
I(E_{\gamma}) &=& \int_{0}^{z_{\textrm{max}}}dz \int_{\Gamma_{\textrm{min}}}^{\Gamma_{\textrm{max}}} d\Gamma
\int_{L_{\gamma}^{\textrm{min}}}^{L_{\gamma}^{\textrm{max}}} \frac{dV}{dz} \nonumber \\
&\times& \Phi (L_{\gamma}, \Gamma, z) \cdot \frac{dN}{dE}(E_{\gamma}(1+z)) \cdot \frac{L_{\gamma}}{2 \pi (d_{L}(z))^{2}} \nonumber \\
&\times& exp \left[ - \tau(z, E_{\gamma}(1+z)) \right], 
%\cdot (1-\omega_{\gamma}(L_{\gamma},z)),
\label{eq:I_AGN}
\end{eqnarray}
where $E_{\gamma}$ is the observed energy of the gamma rays. $\Gamma$ is taken to be within $[1.6, 2.4]$. The maximum redshift we 
assume is $z_{\textrm{max}}=4$. We integrate over sources with bolometric luminosities 
from $0.7 \times 10^{44}$ to $10^{50}$ erg/s. $\frac{dN}{dE}(E_{\gamma}) \propto E_{\gamma}^{-\Gamma}$ is the averaged spectrum 
from these sources. 
When comparing to the isotropic gamma-ray background an additional multiplication term of $(1-\omega_{\gamma}(L_{\gamma},z))$ is included in Eq.~\ref{eq:I_AGN}, with $\omega_{\gamma}$ is the fraction of sources that are below the detection threshold.
For the BL Lacs and FSRQs that contribute to the \textit{Fermi} isotropic background, Ref.~\cite{Cholis:2024hmd} took $\omega_{\gamma}=0.9$.
However, for the EGRB data from \textit{COMPTEL}, \textit{EGRET} and \textit{Fermi}, we use $\omega_{\gamma}=0$.

We also note that at energies below $\simeq 50$ MeV, the expected spectrum from BL Lacs and FSRQs will have a different slope as has been shown in specific example sources \cite{Collmar:2010qd, MAGIC:2024lrb}. We take for both BL Lacs and FSRQs to have a spectral index of $\Gamma_{E<50 \textrm{MeV}} = 1.5 \pm 0.2$ which we marginalize within the quoted uncertainty in our fits. 

\begin{figure}
\begin{centering}
\hspace{-0.0cm}
\includegraphics[width=3.6in,angle=0]{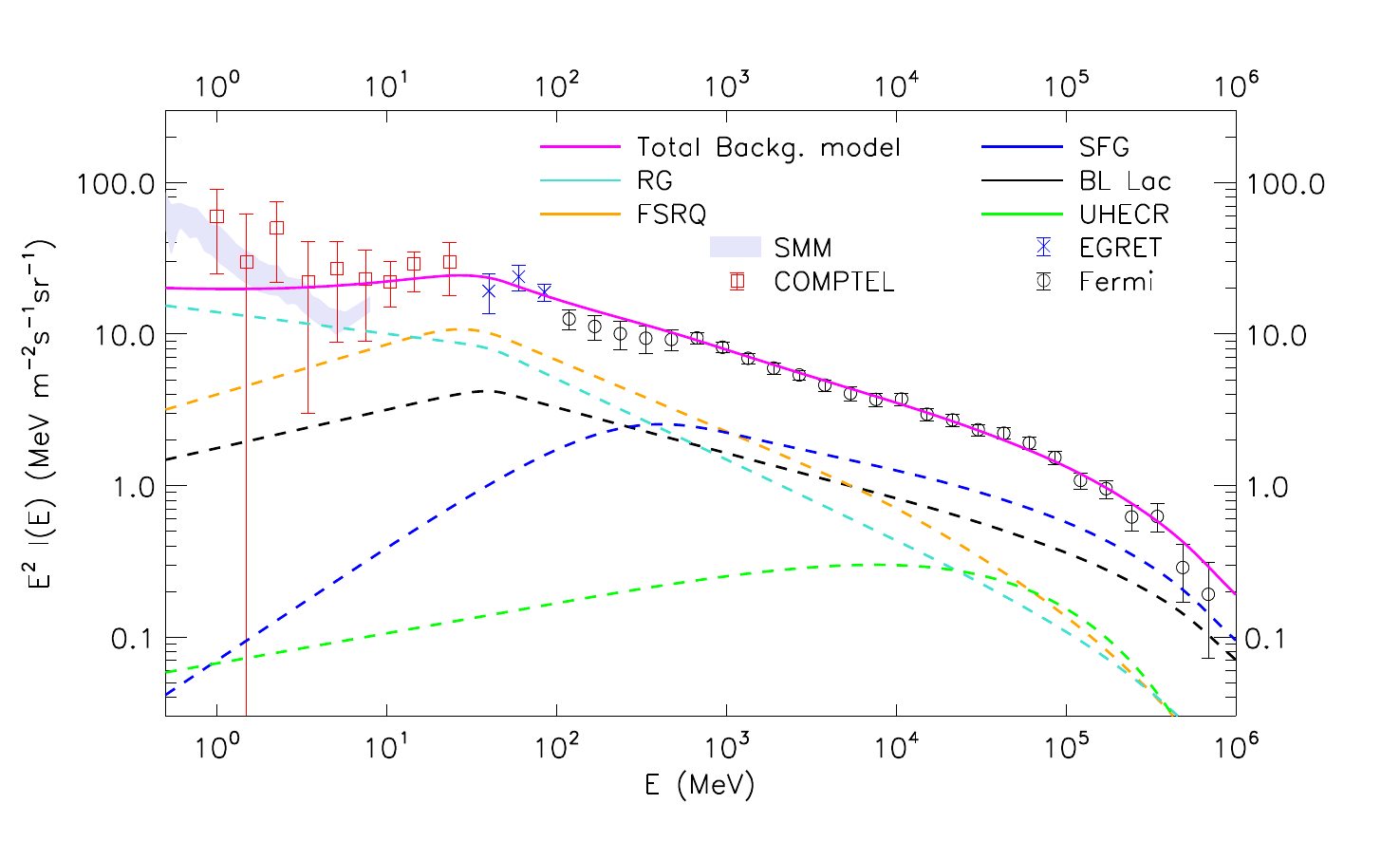}
\end{centering}
\vspace{-0.6cm}
\caption{
Best-fit model to the \textit{Fermi} EGRB spectrum (model A) of Ref.~\cite{Fermi-LAT:2014ryh}. We assume a combination of radio 
galaxies (turquoise dashed line), star-forming and starburst galaxies (blue dashed line), FSRQ sources (orange dashed line), BL Lac sources 
(black dashed line) and UHECRs (green dashed line). 
The combined fit (magenta solid line) gives a best fit of $\chi^{2}$/dof=0.96, to the combined \textit{Fermi}, \textit{COMPTEL} and \textit{EGRET} data. The 
measurement from \textit{SMM} is given only for comparison (see text for further details).}
\label{fig:BackgroundModel}
\end{figure}

In Fig.~\ref{fig:BackgroundModel}, we fit to the \textit{Fermi}, \textit{COMPTEL} and \textit{EGRET} EGRB spectrum the combination of BL Lacs, FSRQs, radio galaxies (RG), star-forming galaxies (SFG)
and UHECR-produced gamma rays.
The normalization and spectral index of each component have been allowed to vary in that fit. 
We note that in Fig.~\ref{fig:BackgroundModel}, we also show the \textit{SMM} data for comparison. However, since these are not peer-reviewed data, we avoid including them in the fit as they would dominate the derived PBH limits. 

\vspace{-0.3cm}
\subsection{Star-forming and Starburst Galaxies}
\label{subsec:Starforming}

The combined contribution of star-forming and starburst galaxies to the EGRB is given by a similar formula to that 
of Eq.~\ref{eq:I_AGN}. However, there are some important differences. First, only a very small number of such galaxies 
have been identified in gamma rays if one excludes the emission from the AGN at their cores. 
%Thus, for this component we set $\omega_{\gamma}=0$ in Eq.~\ref{eq:I_AGN} when fitting to any of the IGRB spectra. 
Thus, for this component its EGRB contribution is nearly equal to its isotropic background contribution.  
Furthermore, we take 
$L_{\gamma}$ in the range of $10^{37}-10^{41}$ erg/s and integrate up to $z_{\textrm{max}}=5$. 

While most of these galaxies have not been identified at gamma-ray energies many more have been 
detected in the infrared. Therefore, we can use the infrared observations to build a bolometric luminosity
function for gamma rays, using Ref.~\cite{2012ApJ...755..164A}, that correlates the infrared and gamma-ray 
bolometric luminosities of star-forming galaxies,  
\begin{equation}
log_{10}\left( \frac{L_{\gamma}}{\textrm{erg s}^{-1}}\right) = \alpha \cdot log_{10} \left( \frac{L_{\textrm{IR}}}{10^{10} L_{\odot}}\right) + \beta.
\label{eq:LIR_to_Lgamma}
\end{equation}
We take $\alpha = 1.17$ and $\beta = 39.28$~\cite{2012ApJ...755..164A}.

Infrared observations have identified three sub-populations of star-forming galaxies that follow different luminosity functions
~\cite{Gruppioni:2013jna}.  
These galaxy populations are regular star-forming galaxies (SF), star-forming galaxies that contain an AGN (SF-AGN) and starburst (SB) galaxies.
Their combined luminosity function is
\begin{eqnarray}
\Phi_{\gamma} (L_{\gamma}, z, f_{sb}) &=& (1 - f_{sb}) \cdot [\Phi_{\textrm{IR}}^{\textrm{SF}}(L_{\textrm{IR}}(L_{\gamma}),z,\{g^{\textrm{SF}}\}) \nonumber \\
&+& \Phi_{\textrm{IR}}^{\textrm{SF-AGN}}(L_{\textrm{IR}}(L_{\gamma}),z,\{g^{\textrm{SF-AGN}}\}) ] \nonumber \\
&+& f_{sb} \cdot \Phi_{\textrm{IR}}^{\textrm{SB}}(L_{\textrm{IR}}(L_{\gamma}),z,\{g^{\textrm{SB}}\}).
\label{eq:SF_GLF}
\end{eqnarray}
The fraction in the luminosity function that is contributed by the SB galaxies, $f_{sb}$ is taken to be 0.5.
The IR luminosity functions of SF galaxies ($\Phi_{\textrm{IR}}^{\textrm{SF}}$), of SF-AGN galaxies ($\Phi_{\textrm{IR}}^{\textrm{SF-AGN}}$) 
and of SB galaxies ($\Phi_{\textrm{IR}}^{\textrm{SB}}$), are parametrized by $\{g^{\textrm{SF}}\}$, $\{g^{\textrm{SF-AGN}}\}$ and $\{g^{\textrm{SB}}\}$ respectively~\cite{Gruppioni:2013jna}.

While there are expected to be significant variations in the gamma-ray spectra between different galaxies, their averaged spectrum is,
\begin{eqnarray}
\frac{dN}{dE} \propto && \left( \frac{E}{\textrm{1 GeV}} \right)^{-\gamma_{1}} \cdot \left(1 - exp\left[ \frac{-E}{E_{0}} \right] \right) \nonumber \\
&+&  \left( \frac{E}{\textrm{1 GeV}} \right)^{-\gamma_{2}} \cdot exp\left[ \frac{-E}{E_{0}} \right].
\label{eq:SFG_Spectrum}
\end{eqnarray}
There is no integration over a range of spectral indices as $\Gamma$ of Eq.~\ref{eq:I_AGN}. Instead, we model the 
diffuse emission in those galaxies with $\gamma_{1}=2.2$, $\gamma_{2} = 1.2-1.8$ and $E_{0} = 0.3$ GeV. In the fitting 
we allow for some small variation on the high-energy spectral index $\gamma_{1}$. As with the case of BL Lacs and FSRQs, SFGs will have a different spectral index at $E\lesssim 300$ MeV, where the $\pi^{0}$ component is taken over by the ICS \cite{Lacki:2012si, Chakraborty:2012sh, Tamborra:2014xia}. This is described by $\gamma_{2}$. In our fits, we marginalize over the value of $1.2 \leq \gamma_{2} \leq 1.8$. In Fig.~\ref{fig:BackgroundModel}, we show an example spectrum 
from SFGs.

\vspace{-0.3cm}
\subsection{Modeling the contribution from Radio Galaxies}
\label{subsec:Radio}

Radio galaxies (RG), classified as Fanaroff-Riley (FR) type I and II, emit most of their gamma rays 
from misaligned relativistic jets originating from the central black hole \cite{Urry:1995mg}. 
While only a small number of these galaxies have been detected by the \textit{Fermi}-LAT
in gamma rays \cite{Fermi-LAT:2019yla}, their contribution to the EGRB is expected to be a significant one 
(see e.g. \cite{2011ApJ...733...66I, Stecker:2010di, Inoue:2011bm, DiMauro:2013xta, Hooper:2016gjy, 
Fukazawa:2022gwm, Cholis:2024hmd}). Like with star-forming galaxies, to model their emission spectra and 
luminosity distribution in gamma rays we use observations from lower energy photons; in this case
radio waves. The bolometric luminosity of RGs in gamma rays between 0.1 and 
100 GeV $L_{\gamma, \, \textrm{RG}}$, can be correlated to the luminosity from their core at 5 GHz $L_{\textrm{5 GHz}}$
\cite{DiMauro:2013xta, Hooper:2016gjy, Stecker:2019ybn}.
We follow the work of Ref.~\cite{Blanco:2021icw}, that gives the probability for a radio galaxy to have $L_{\gamma, \, \textrm{RG}}$
for a given $L_{\textrm{5 GHz}}$,
\begin{eqnarray}
&&P(L_{\gamma, \, \textrm{RG}}, L_{\textrm{5 GHz}}) = \frac{1}{\sqrt{2 \pi \sigma_{\textrm{RG}}^2}} \\
&\times& exp \left[ - \frac{\left( log_{10}\left( \frac{L_{\gamma, \, \textrm{RG}}/(\textrm{1 erg s}^{-1})}{ (L_{\textrm{5 GHz}}/(10^{40} \textrm{ erg s}^{-1}))^{b_{\textrm{RG}}} }\right) -d_{\textrm{RG}} \right)^2}{2 \sigma_{\textrm{RG}}^2} \right]. \nonumber
\label{eq:Prop_Lgamma}
\end{eqnarray}
This relation accounts for the fact that there is a wide scatter in the correlation between $L_{\gamma, \, \textrm{RG}}$
and $L_{\textrm{5 GHz}}$. We take $b_{\textrm{RG}} = 0.78$, $d_{\textrm{RG}} = 40.78$ and $\sigma_{\textrm{RG}} = 0.88$ \cite{Blanco:2021icw}.

The luminosity function of RGs in gamma rays $\Phi_{\gamma}$ is related to the luminosity function from their cores at 5 GHz, $\Phi_{\textrm{RG c}}$
through \cite{Blanco:2021icw}, 
\begin{eqnarray}
\Phi_{\gamma}(L_{\gamma}, z) = \frac{1}{b_{\textrm{RG}}} \int_{L_{\textrm{5 GHz}}^{\textrm{min}}}^{L_{\textrm{5 GHz}}^{\textrm{max}}}
&& \frac{d L_{\textrm{5 GHz}}}{L_{\textrm{5 GHz}}} \cdot \Phi_{\textrm{RG c}}(L_{\textrm{5 GHz}}, z) \nonumber \\
&\times& P(L_{\gamma}, L_{\textrm{5 GHz}}).
\label{eq:Phi_gamma}
\end{eqnarray}
We use as integration limits $L_{\textrm{5 GHz}}^{\textrm{min}} = 10^{37}  \textrm{erg s}^{-1}$ and 
$L_{\textrm{5 GHz}}^{\textrm{max}} = 10^{43}  \textrm{erg s}^{-1}$. For $\Phi_{\textrm{RG c}}(L_{\textrm{5 GHz}}, z)$ 
we use the parameterizations of Ref.~\cite{Blanco:2021icw} and~\cite{2018ApJS..239...33Y} (see also \cite{Cholis:2024hmd}). 

The spectrum of gamma rays with energy above 50 MeV from these galaxies is taken to be on average,
\begin{equation}
\frac{dN}{dE} \propto \left( \frac{E}{\textrm{1 GeV}} \right)^{-2.39}.
\label{eq:RG_Spectrum}
\end{equation}
As with the other sources, we allow in the fit for some degree of freedom on the high-energy spectral index (see more discussion in section~\ref{subsec:Combination}). At energies of $E \lesssim 50$ MeV, we take the spectrum of RGs to have instead a value of $\Gamma_{E \leq 50 \textrm{MeV}}^{RG} = 1.6 \pm 0.4$. The change in the spectrum of RGs at lower energies is based on our expectation of a changing mechanism in their production (low-energy ICS) and on a small number of observations \cite{Kino:2006js, Kino:2013tza, Muller:2016xwh, Gan:2021nzt, Gan:2024fbf}. We marginalize in our fits the $\Gamma_{E \leq 50 \textrm{MeV}}^{RG}$ within the quoted range. 

The observed intensity of gamma rays from radio galaxies is,
\begin{eqnarray}
I(E_{\gamma}) &=& \int_{0}^{z_{\textrm{max}}}dz 
\int_{L_{\gamma}^{\textrm{min}}}^{L_{\gamma}^{\textrm{max}}} \frac{dV}{dz}  \Phi_{\gamma} (L_{\gamma}, z) \cdot \frac{dN}{dE}(E_{\gamma}(1+z)) \nonumber \\
&\times& \frac{L_{\gamma}}{2 \pi (d_{L}(z))^{2}} \cdot exp \left[ - \tau(z, E_{\gamma}(1+z)) \right].
\label{eq:I_RG}
\end{eqnarray}
We integrate up to $z_{\textrm{max}}=5$, for bolometric luminosities between $10^{38}$ and $10^{42.5}$
erg/s.  
In Fig.~\ref{fig:BackgroundModel}, we show an example spectrum 
from RGs fitted in combination with other sources to the \textit{Fermi}-LAT, \textit{EGRET} and \textit{COMPTEL} EGRB measurements.

\vspace{-0.3cm}
\subsection{Modeling the contribution from Ultra-High-Energy Cosmic Rays}
\label{subsec:UHECRs}

Gamma rays can also be produced as part of the cascade products from inelastic scattering events 
of UHECRs with the cosmic microwave background \cite{1966PhRvL..16..748G, Zatsepin:1966jv} or even 
later on from the inverse Compton scattering with low-energy photons of high-energy electrons and positrons 
produced in those cascades. That flux of gamma rays also gets attenuated.
The resulting spectrum is, 
\begin{equation}
\frac{dN}{dE} = A_{\textrm{UHE}} \left( \frac{E}{\textrm{1 GeV}} \right)^{-\gamma_{\textrm{UHE}}} exp\left[ -\left( \frac{E}{E_{\textrm{cut}}}\right)^{\beta_{\textrm{UHE}}} \right].
\label{eq:UHECR_to_gamma_ray_Spectrum}
\end{equation}
Following Refs.~\cite{Cholis:2013ena, Cholis:2024hmd}.
we take $A_{\textrm{UHE}} = 3.0 \times 10^{-8}$ $\textrm{GeV}^{-1} \textrm{cm}^{-2} \textrm{s}^{-1} \textrm{sr}^{-1}$, $\gamma_{\textrm{UHE}} = 1.78$, $\beta_{\textrm{UHE}} = 0.54$ and $E_{\textrm{cut}} = 40$ GeV. 

To account for uncertainties in the chemical composition of UHECRs, in the redshift distribution of the UHECR sources, 
in the infrared background and in intergalactic magnetic fields we allow in our fits to the EGRB data for a large range 
on the normalization of the gamma-ray flux  from the original expectation of Eq.~\ref{eq:UHECR_to_gamma_ray_Spectrum}. 
An example of the spectrum is shown in Fig.~\ref{fig:BackgroundModel}. 

\vspace{-0.3cm}
\subsection{The MeV-scale spectrum from radiating Primordial Black Holes}
\label{subsec:DM}

Evaporating PBHs give gamma rays at MeV energies through four different mechanisms. There are gamma rays as a result of Hawking radiation whose energy is directly related to the PBH's mass. This is known as the primary component and is responsible for the highest energy gamma rays produced by PBHs of a given mass. However, as their mass is decreasing and subsequently their temperature is increasing, PBHs can emit significant amounts of energy in massive fundamental particles that are unstable and will undergo hadronization and decays which also give gamma rays. This second mechanism of producing gamma rays from Hawking radiation is known as secondary emission and gives wide spectra. Furthermore, there is final state radiation from the higher energy particles produced by the PBHs, giving additional MeV-range gamma rays. Finally, as PBHs produce high fluxes of electrons and positrons both directly and as a result of unstable species hadronizing and decaying, we need to account for the fact that the produced positrons may undergo pair annihilation as they propagate though their local interstellar medium. This last component is known as in-flight annihilation. 

\begin{figure}
\begin{centering}
\hspace{-0.0cm}
\includegraphics[width=3.6in,angle=0]{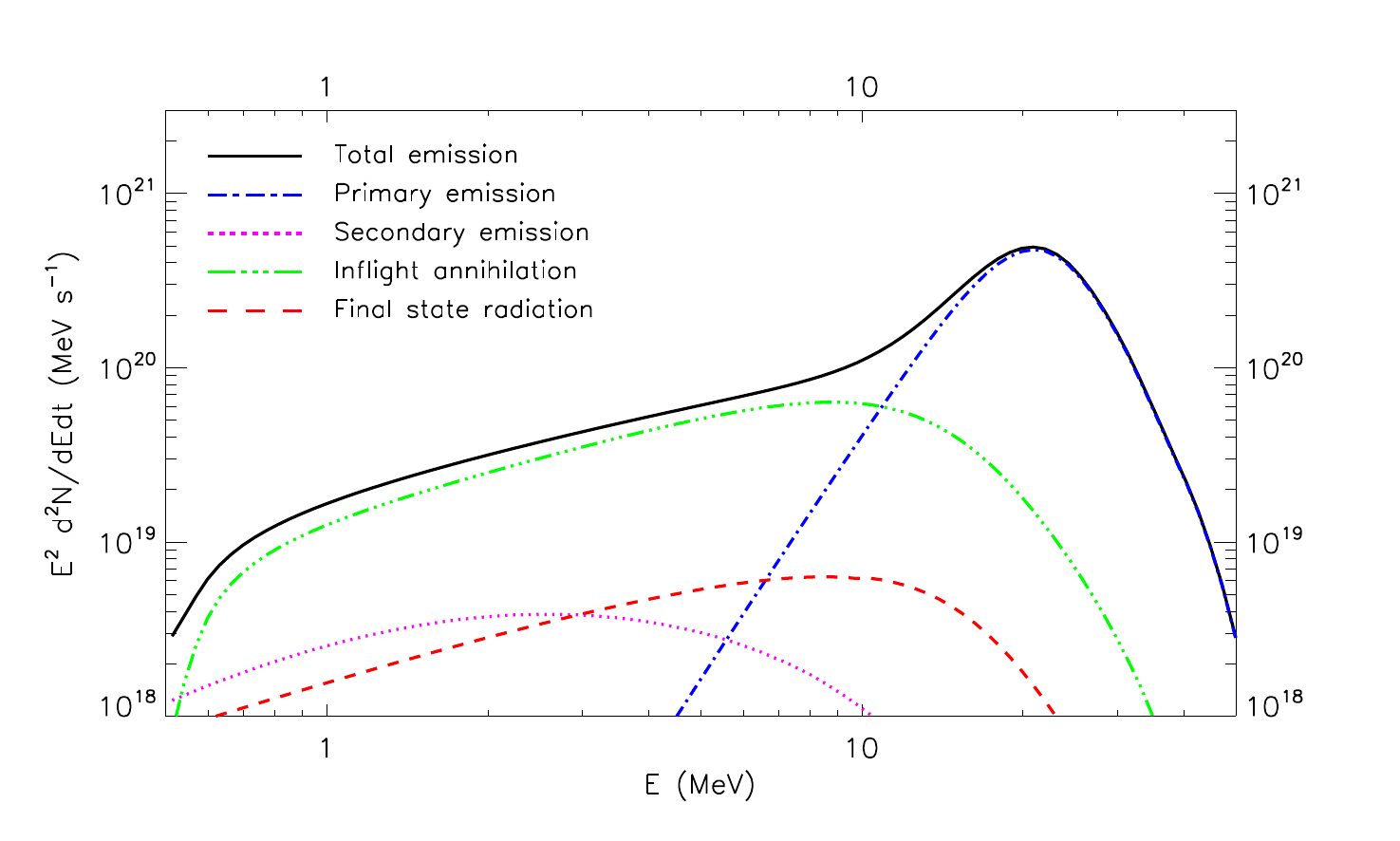}
\end{centering}
\vspace{-0.7cm}
\caption{
The gamma-ray spectrum produced by a PBH of mass $m_{\textrm{PBH}} = 3.2 \times 10^{15}$ g at its own (source) frame. With the blue dash-dotted line we show the primary/direct Hawking component that dominates the highest energies and is responsible for most of the power in gamma rays from the evaporation of PBHs. The secondary (purple dotted line) and the final state radiation (red dashed line) contribute only a few $\%$ of the total power in gamma rays. However, the in-flight annihilation component (green dash-triple-dotted line) is the dominant component at energies $\lesssim 0.5$ the peak energy. The total power is given in the black line (see also text for details).}
\vspace{-0.6cm}
\label{fig:PBHcomponents}
\end{figure}

In Fig.~\ref{fig:PBHcomponents}, for a single PBH of mass $m_{\textrm{PBH}} = 3.2 \times 10^{15}$ g, we show the resulting differential gamma-ray spectra per unit time from these four components multiplied by $E^{2}$, i.e. the $E^{2} \, d^{2}N/dE dt$. That allows us to see the gamma-ray energy where the power from each component peaks. As we show, the primary component (dash-dotted blue line) is the most important component at the highest energies, while the other three components and most importantly among them the in-flight annihilation component (dash-triple-dotted green line), dominate the emission at energies below the peak at $\simeq 20$ MeV for such a PBH. 
As we discuss in Section~\ref{sec:PBHlimits}, including the contribution of these lower energy gamma rays is important in setting accurate limits on the abundance of PBHs and in making projections on the sensitivity of future gamma-ray detectors to search for PBHs. 

To evaluate these four components we use \texttt{GammaPBHPlotter} of Ref.~\cite{Carlini:2025bki}, which also takes as input the primary and secondary gamma-ray components from \texttt{BlackHawk} \cite{Arbey:2021mbl}. \texttt{GammaPBHPlotter} \cite{Carlini:2025bki, Carlini:2025bkiZenodo} is an open-source software developed to quickly and accurately simulate the gamma-ray Hawking spectra of PBHs in the mass range of $10^{14}$-$10^{19}$ g. This tool can be used to generate the gamma-ray spectra from each of the previously described four components, as well as their sum and thus provide an accurate evaluation of the gamma-ray spectrum generated from a PBH of a particular mass. 
We evaluate spectra from a list 
of 45 unique masses $m_{\textrm{PBH}}$ from $1 \times 10^{14}$ to  $3.1 \times 10^{17}$ g. Those values are for a PBH's mass at $z=0$. 

The contribution to the EGRB's intensity at energy $E$, formally should only come from PBHs outside the Milky Way. However, given that all instruments use the high galactic latitudes to derive the EGRB spectrum, if dark matter PBHs give a gamma-ray contribution one has to include the contribution from PBHs in the Milky Way's high latitudes, i.e. the Galactic component $I_{\textrm{MW}}(E)$. That component is to be added to the contribution from all the PBHs in all other galaxies, i.e. the Extragalactic component $I_{\textrm{ExGal}}(E)$,
\begin{equation}
I(E) = I_{\textrm{MW}}(E) + I_{\textrm{ExGal}}(E).
\label{eq:Total_Intensity}
\end{equation} 
The $I_{\textrm{MW}}(E)$ depends on the exact region of interest (region on the sky) each satellite uses to determine the EGRB spectrum and is set by integrating along the line of sight $l.o.s.$ within that region of interest,  
\begin{equation}
I_{\textrm{MW}}(E) =  \frac{d^{2}N}{dE dt}(E) \frac1{\Omega_{\textrm{ROI}}}  \int_{l.o.s.} d \ell ~ \frac{ \rho_{\textrm{DM}}(\ell)}{m_{\textrm{PBH}}}. 
\label{eq:MW_Intensity}
\end{equation} 
$\Omega_{\textrm{ROI}}$ is the solid angle contained within the region of interest, $\rho_{\textrm{DM}}(\ell)$ the dark matter density of the Milky Way at position $\ell$ away from our location in the Galaxy. 
If chosen correctly that region of interest gives on average the smallest contribution from the Milky Way to the intensity of Eq.~\ref{eq:MW_Intensity}. 

For the extragalactic component, its intensity is given by integrating over redshift $z$, 
\begin{eqnarray}
I_{\textrm{ExGal}}(E) &=& \frac{1}{4\pi} \int_{0}^{z_{\textrm{max}}} dz \,  \frac{c}{H(z)} \frac{d^{2}N}{dE dt}( (1+z)E, z) \nonumber \\
&\times& \left( \frac{\Omega_{dm} \rho_{c}}{m_{\textrm{PBH}}} \right) \, exp \left[ - \tau(z, E) \right] .  
\label{eq:EGRB_DM_Intensity}
\end{eqnarray} 
$H(z) = H_{0} \sqrt{  \Omega_{\Lambda} + \Omega_{k}(1+z)^2 + \Omega_{m} (1+z)^{3} + \Omega_{r}(1+z)^4 }$ 
with $H_{0} = 100 h$ the values for the Hubble expansion scaling factor $h$, and the dark energy density parameter $\Omega_{\Lambda}$, 
matter density parameter $\Omega_{m}$, radiation density parameter $\Omega_{r}$ and curvature parameter 
$\Omega_{k}$, taken from the \textit{Planck}-Collaboration results \cite{Planck:2018vyg}.
$\Omega_{dm}$ is the current dark matter density parameter and
$\rho_{c}$ the present critical density. The differential spectrum per unit time $d^{2}N/dEdt$ not only needs to be redshifted by the $1+z$ factor from its source, 
but its shape and amplitude also change with redshift as a PBH of mass 
$m_{\textrm{PBH}}$ at $z=0$ would have been more massive at earlier times and produced at the source frame a smaller amount of power and at lower energies compared to later times (again at the source frame). Since we focus on PBHs with masses that can be constrained by the MeV gamma-ray observations, our mass range of study does include PBHs whose mass has changed by a significant amount from our maximum redshift of integration $z_{\textrm{max}} = 10$ in Eq.~\ref{eq:EGRB_DM_Intensity}, to our current era of $z=0$. 
For each of the 45 $m_{\textrm{PBH}}$ values at $z=0$, we backward evolve the mass and spectrum at the source frame before redshifting the entire spectrum \footnote{We note that even for the smallest value of $m_{\textrm{PBH}}$ that we test, i.e. $ m_{\textrm{PBH}}(z=0) = 1.0 \times 10^{14} $ g, the mass evolution of the PBHs is up to a factor of 2. If all dark matter was in such mass PBHs that would result in the abundance of dark matter to drop by a factor of 2 from $z=10$ to $z=0$; which would be in tension with our cosmological observations. However, as we show such PBHs can only account for up to $10^{-10}$ of the dark matter abundance.}. 
We use \texttt{GammaPBHPlotter} to produce the emitted spectra at 101 linearly spaced redshift values from $z=10$ to $z=0$. The exponential suppression in Eq.~\ref{eq:EGRB_DM_Intensity}, accounts for the attenuation of gamma rays by pair production with UV photons along their path from their source to the detectors. However, for the PBH masses of interest, their maximum gamma-ray energy at the source frame is only as high as a few GeV; where this attenuation factor is insignificant i.e. $exp \left[ - \tau(z, E) \right] = 1$ \footnote{As we described in Sections~\ref{subsec:BLLac-FSRQ}-\ref{subsec:Radio}, we account for the gamma-ray attenuation of the background components as those are modeled to energies as high as 1 TeV at which that suppression is important.}.  

We note that even in Eq.~\ref{eq:EGRB_DM_Intensity}, by using the EGRB data we do not need to remove the PBH emission from the galaxies that are identified as individual extragalactic sources in MeV gamma rays. 
Since most of the emission to the dark matter EGRB spectrum comes from the extragalactic component,
for simplicity we take that the PBH intensity contribution to the EGRB is,
\begin{equation}
I(E) \simeq I_{\textrm{ExGal}}(E),
\label{eq:Total_Intensity_Approx}
\end{equation} 
which we estimate to be accurate at the $10\%$ level. 

\begin{figure}
\begin{centering}
\hspace{-0.0cm}
\includegraphics[width=3.6in,angle=0]{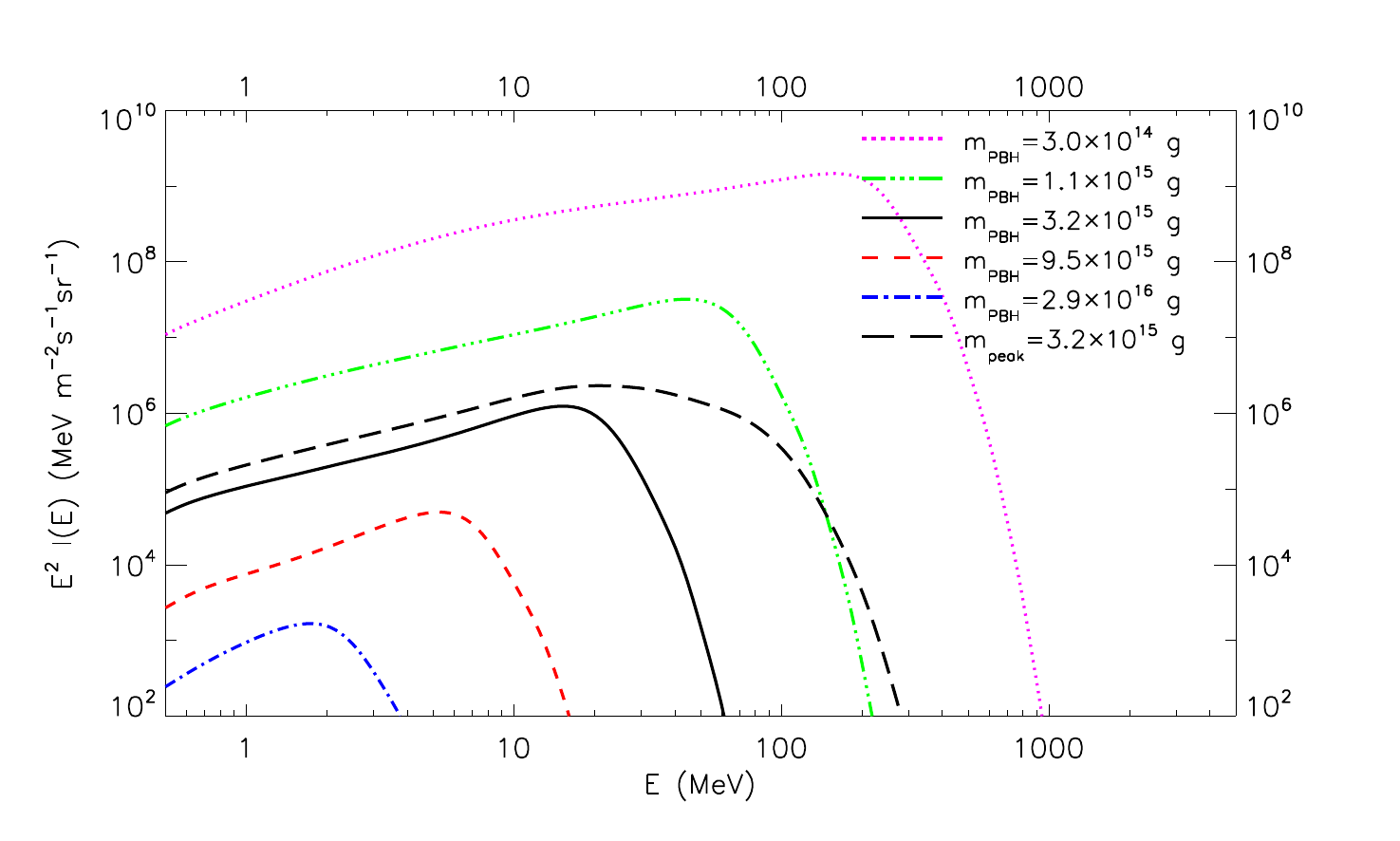}
\end{centering}
\vspace{-0.7cm}
\caption{
The dark matter contribution to the EGRB from PBHs of different monochromatic masses $m_{\textrm{PBH}}$ in the range of $3.0 \times 10^{14}$-$2.9 \times 10^{16}$ g. MeV-scale gamma-ray observations probe that PBH mass range. PBHs with mass $\gtrsim 3 \times 10^{17}$ g have Hawking emission that gives a weak signal  and at energies not probed by MeV-scale detectors, while PBHs with mass $\lesssim 10^{14}$ g have already evaporated. In the black long-dashed line we also give an example of the EGRB spectrum from an extended PBH mass-distribution with a peak mass at formation of $m_{\textrm{peak}} = 3.2 \times 10^{15}$ g.}
\label{fig:AltPBHtotalSpectra}
\end{figure}

In Fig.~\ref{fig:AltPBHtotalSpectra}, we give the expected $E^{2} \, I(E)$ for five different monochromatic PBH masses at $z=0$ with values between $3.0 \times 10^{14}$ and $2.9 \times 10^{16}$ g. 
The black solid line is evaluated for a $m_{\textrm{PBH}}(z=0) = 3.2 \times 10^{15}$ g, i.e. the same mass for which in Fig.~\ref{fig:PBHcomponents}, we give the $E^{2} d^{2}N/dEdt$ at the source frame.
Properly accounting for the evolution of the PBH's mass and gamma-ray spectrum at its source frame, and also properly integrating over the combined spectra of PBHs from many redshifts, has the effect 
of making less prominent the spectral peak of the PBH contribution to the EGRB. Moreover, since we do include the important in-flight annihilation component our PBH spectra $E^{2} \, I(E)$, 
are enhanced by a factor of $\gtrsim 5$ at energies that are $\lesssim 0.5$ the peak energy of the primary/direct Hawking component (see e.g. \cite{Xie:2024eug}). 

In addition to studying alternative choices for monochromatic PBH masses, we test the more realistic case of PBHs having an extended distribution of masses as a result of a distribution of primordial curvature fluctuations.

Following the prescription of Ref.~\cite{Biagetti:2021eep}, we generate primordial curvature perturbations that follow a non-Gaussian distribution (but still relatively close to a Gaussian one).  
Subsequently, those perturbations create an extended mass distribution that at formation has a peak $m_{\textrm{peak}}$ (see Fig. 3 of Ref.~\cite{Biagetti:2021eep}). 
The width of the non-Gaussian PBH mass distribution is set by the $m_{\textrm{peak}}$ value and scales proportionally to it. Thus, such a distribution is defined by that one parameter. While there is some possible additional freedom in the parameterization of the PBH's masses and the width of that distribution can be used as an additional parameter, Ref.~\cite{Biagetti:2021eep}, makes a case for why a single parameter can describe the entire population of PBHs' masses. For simplicity we follow that assumption. 
We have created 31 alternative choices of extended mass distributions with values for $m_{\textrm{peak}}$ between $1.3 \times 10^{15}$ and $3.1 \times 10^{17}$ g (logarithmically spaced). 

Similarly to the monochromatic case, we evolve the masses of the PBHs as those evaporate with time, with the smaller masses of a given distribution, evolving the fastest. As we describe in our results section, the lowest of these choices for the extended mass distribution are heavily suppressed.
In Fig.~\ref{fig:AltPBHtotalSpectra}, we also present the resulting EGRB spectrum from an extended PBH mass distribution with a peak mass at formation \footnote{For simplicity we take the PBH formation redshift to be at matter-radiation equality i.e. $z=3400$.} of $m_{\textrm{peak}} = 3.2 \times 10^{15}$ g. 

\vspace{-0.3cm}
\subsection{Combining the gamma-ray source classes to explain the EGRB}
\label{subsec:Combination}

Given that we want to derive limits on PBH dark matter, we first fit the combined \textit{Fermi}, \textit{EGRET} and \textit{COMPTEL} EGRB spectrum to the combination
of astrophysical sources described without including any potential PBH-originated flux.
Each of those components has a relevant freedom
in its normalization and other than the gamma-ray component from the UHECRs all other components have freedom also in their high-energy spectral index by $\Delta \gamma$.  
For the RGs, the BL Lacs and the FSRQs the high-energy spectral index is defined for energies above 50 MeV, while for the SFGs it is above the $\pi^{0}$ bump i.e. above $\simeq 3\times 10^2$ MeV. 
For the combined spectrum from BL Lacs and the combined spectrum from FSRQs we take the low-energy spectral index to be between 1.3 and 1.7 and marginalize over that range in the fit separately for each population. We also marginalize over the low-energy spectral index of radio galaxies  $1.2 \leq \Gamma_{E \leq 50 \textrm{MeV}}^{RG} \leq 2.0$ and over the low-energy spectral index of SFGs $1.2 \leq \gamma_{2} \leq 1.8$ (see sections\ref{subsec:Starforming} and~\ref{subsec:Radio}).

In Table~\ref{tab:FitFreedom}, we give the relevant ranges on the normalization of each component and the range of spectral indices change $\Delta \gamma$ that we allow in our fits. 
These ranges are chosen to account generously for the relevant modeling uncertainties and have been previously tested in Ref.~\cite{Cholis:2024hmd}. 
Our limits on the PBH abundance presented in Section~\ref{sec:PBHlimits}, are practically unaffected by the exact ranges when marginalizing over the background components' normalizations and spectral indices.  
\begin{table*}[t]
    \begin{tabular}{cccc}
         \hline
           Component \; &  Normalization \; & High-energy spectral change $\Delta{\gamma}$ \; & Low-energy \; \\
           &  Range & from reference value & spectral index \; \\
            \hline \hline
            BLLac &  [0.2, 1.0] & [-0.15, +0.15] & [1.3, 1.7]\\   
            FSRQ &  [0.3, 1.0] & [-0.04, +0.03] & [1.3, 1.7]\\
            SF \& SB & [0.7, 1.5] & [-0.3, +0.3] & [1.2, 1.8] \\
            RG &  [0.7, 1.5] & [-0.3, +0.3] & [1.2, 2.0]\\
            UHECR &  [0.3, 3] & 0.0 & NA \\       
        \hline \hline 
        \end{tabular}
\caption{The freedom in normalization and spectral shape of each background astrophysical component when fitting to the EGRB spectrum.} 
\label{tab:FitFreedom}
\end{table*}

In Fig.~\ref{fig:BackgroundModel}, we give the best-fit combination of all background astrophysical sources (magenta solid line) to the combined \textit{Fermi}, \textit{EGRET} and \textit{COMPTEL} EGRB spectrum (spectral model A of Ref.~\cite{Fermi-LAT:2014ryh} for the \textit{Fermi} data). The dashed lines show the contribution from each background component.
Radio galaxies, star-forming galaxies and gamma rays from FSRQs are the three major contributing sources to the EGRB. At energies above 10 GeV BL Lacs and UHECRs are also important. We note that given the modeling freedom of each of the 
components' normalizations and spectra, there can be significant degeneracies as for instance between FSRQs and Radio galaxies. We get a $\chi^{2}$/dof =0.96. 

\vspace{-0.3cm}
\subsection{Statistical analysis}
\label{subsec:analysis}

When we fit the astrophysical/background models to the EGRB spectrum, we minimize the $\chi^2$
loss with respect to the parameters listed in Table~\ref{tab:FitFreedom}. 
When we add an additional contribution from PBHs to the EGRB spectrum, an extra parameter enters our fit, which is the PBH abundance by fraction of dark matter mass in the Universe,
\begin{equation}
f_{\textrm{PBH}} = n_{\textrm{PBH}}\frac{\int dm' \; PDF(m') \cdot m'}{\Omega_{dm} \rho_{c}} \leq 1.
\label{eq:fPBH}
\end{equation} 
The integral is over the PBH mass of a mass-distribution described by a probability density function $PDF(m')$ and $n_{\textrm{PBH}}$ is the number density of PBHs.  
The $f_{\textrm{PBH}}$ parameter acts as a normalization factor for the Hawking radiation flux and reduces the flux from its reference value, evaluated for $f_{\textrm{PBH}} = 1$. 

For our minimization process, we use \path{iminuit} \cite{iminuit,James:1975dr}.
The minimization proceeds as follows: for a given astrophysical background, we first minimize the $\chi^2$ loss with respect to only the astrophysical background parameters ($f_{\textrm{PBH}} = 0$).
Subsequently, we include the PBH component for a specific PBH mass and allow the $f_{\textrm{PBH}}$ to be free between 0 and 1, evaluating upper limits on $f_{\textrm{PBH}}$. 
We apply Wilks' theorem \cite{10.1214/aoms/1177732360}, using the test statistic $LR = -2\log{\Lambda}$, i.e. the difference in $\chi^2$ between the null hypothesis (background-only) and the alternative hypothesis (background plus PBHs).
This test statistic follows a $\chi^2_{\nu}$ distribution, where $\nu$ is the number of additional fitting parameters in the alternative model compared to the null model. In our case, $\nu = 1$ 
\footnote{Given that the null hypothesis with $f_{\textrm{PBH}} = 0$ is at the boundary of the parameter space, the more correct approach is to use Chernoff’s  theorem \cite{10.1214/aoms/1177728725}, in which $LR$ follows a distribution of $\frac{1}{2} \delta(x) + \frac{1}{2}\chi^2$, i.e. the half chi-square distribution, with one degree of freedom \cite{Conrad:2014nna}.}.
We scan the range of $0 \leq f_{\textrm{PBH}} \leq 1$, deriving the $\chi^2$ profile for each PBH mass, and set the limit for $f_{\textrm{PBH}}$ at $\chi^2_{\mathrm{f_{\textrm{PBH}}}} = \chi^2_{\mathrm{f_{\textrm{PBH}}=0}} + 2.71$ which gives us the 95\% upper limits.

\subsection{Improvements upon previous relevant work}
\label{subsec:improvements}

In this subsection, for the ease of the reader, we list the improvements upon previous relevant studies on the limits of PBHs from the extragalactic gamma-ray background. Those are described in full 
detail in the relevant subsections. In short we make four major improvements/contributions.
\begin{enumerate}
\item{We introduce and use a new model for the EGRB spectrum, based on earlier work on the isotropic gamma-ray background in Ref.~\cite{Cholis:2024hmd}.
In our model for the EGRB we calculate the contribution from conventional extragalactic sources including blazars, star-forming galaxies and radio galaxies and also account 
for ultra-high-energy cosmic rays that produce gamma rays when interacting with the infrared background. 
Our EGRB model uses information from the \textit{Fermi} gamma-ray point sources catalog, from  observations at X-rays, the visible spectrum, the infrared and radio waves, and also accounts for modeling uncertainties and variations on the properties within each class of these sources. This model is described in full detail in Sections~\ref{subsec:BLLac-FSRQ}-\ref{subsec:UHECRs},~\ref{subsec:Combination} and~\ref{subsec:analysis}.}
\item{In the evaluation of the gamma-ray signal from the evaporation of the PBHs, we include the fluxes from the final state radiation and the in-flight annihilation components that as we show in Section~\ref{subsec:DM} and Fig.~\ref{fig:PBHcomponents}, are important additions to the gamma-ray flux from the primary and secondary emission of evaporating PBHs, which has been the practice in the literature so far. The final state radiation and the in-flight annihilation components are especially important for the gamma-ray flux at the MeV scale.
The evaluation of all the component fluxes is described in Section~\ref{subsec:DM}.}
\item{We present limits on the PBH abundance not just from monochromatic mass distributions (which is the standard practice), but also from physically motivated extended PBH mass distributions. The modeling of those distributions is given in Section~\ref{subsec:DM}, relying on the use of a newly developed open-source software \texttt{GammaPBHPlotter} of Ref.~\cite{Carlini:2025bki}, developed to rapidly simulate the gamma-ray Hawking spectra of PBHs in the mass range that we probe in our work, shown in Sections~\ref{sec:PBHlimits} and~\ref{sec:Projections}.}
\item{We update the limits on the PBH abundance based on the combination of the above mentioned improvements in Section~\ref{sec:PBHlimits}, and also make projections on the limits of monochromatic and extended mass distributions of PBHs from the future AMEGO-X and e-ASTROGAM telescopes that will be sensitive in the MeV range (described in Section~\ref{sec:Projections}).}

\end{enumerate}
\vspace{-0.4cm}
\section{Current upper limits on the abundance of primordial black holes}
\label{sec:PBHlimits}

\begin{figure}[h]
\begin{centering}
\hspace{-0.0cm}
\includegraphics[width=3.6in,angle=0]{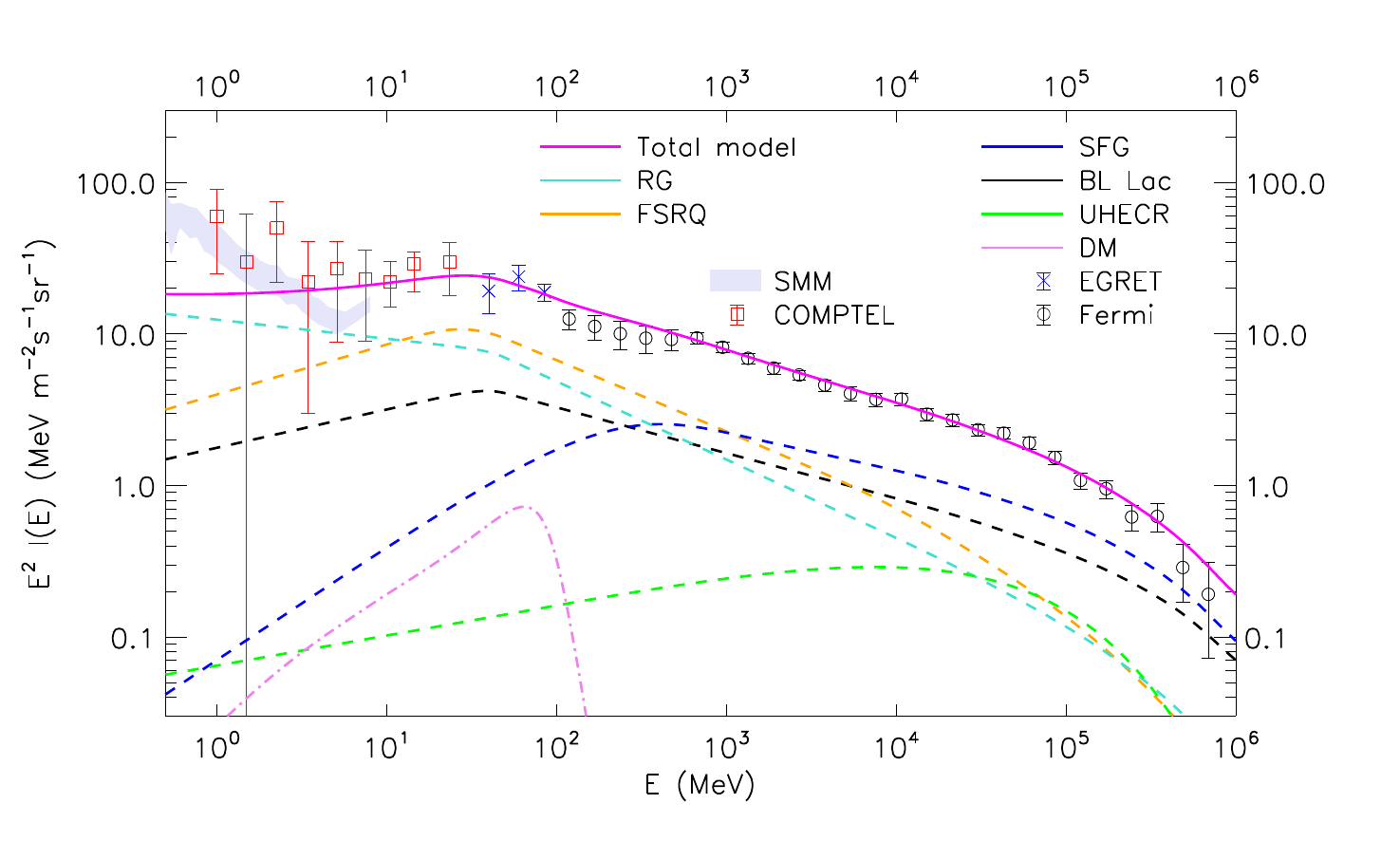}\\
\vspace{-0.3cm}
\includegraphics[width=3.6in,angle=0]{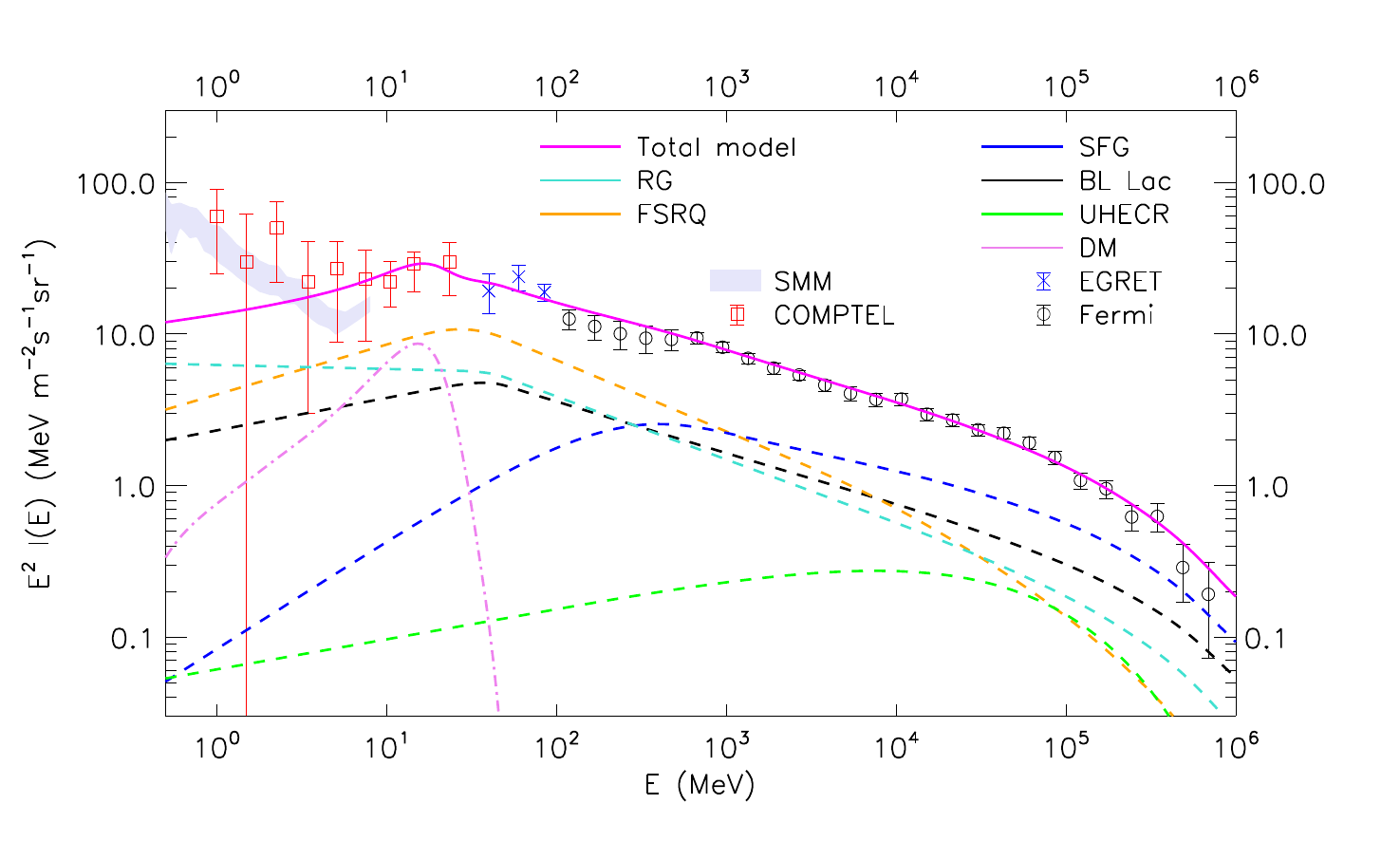}\\
\vspace{-0.3cm}
\includegraphics[width=3.6in,angle=0]{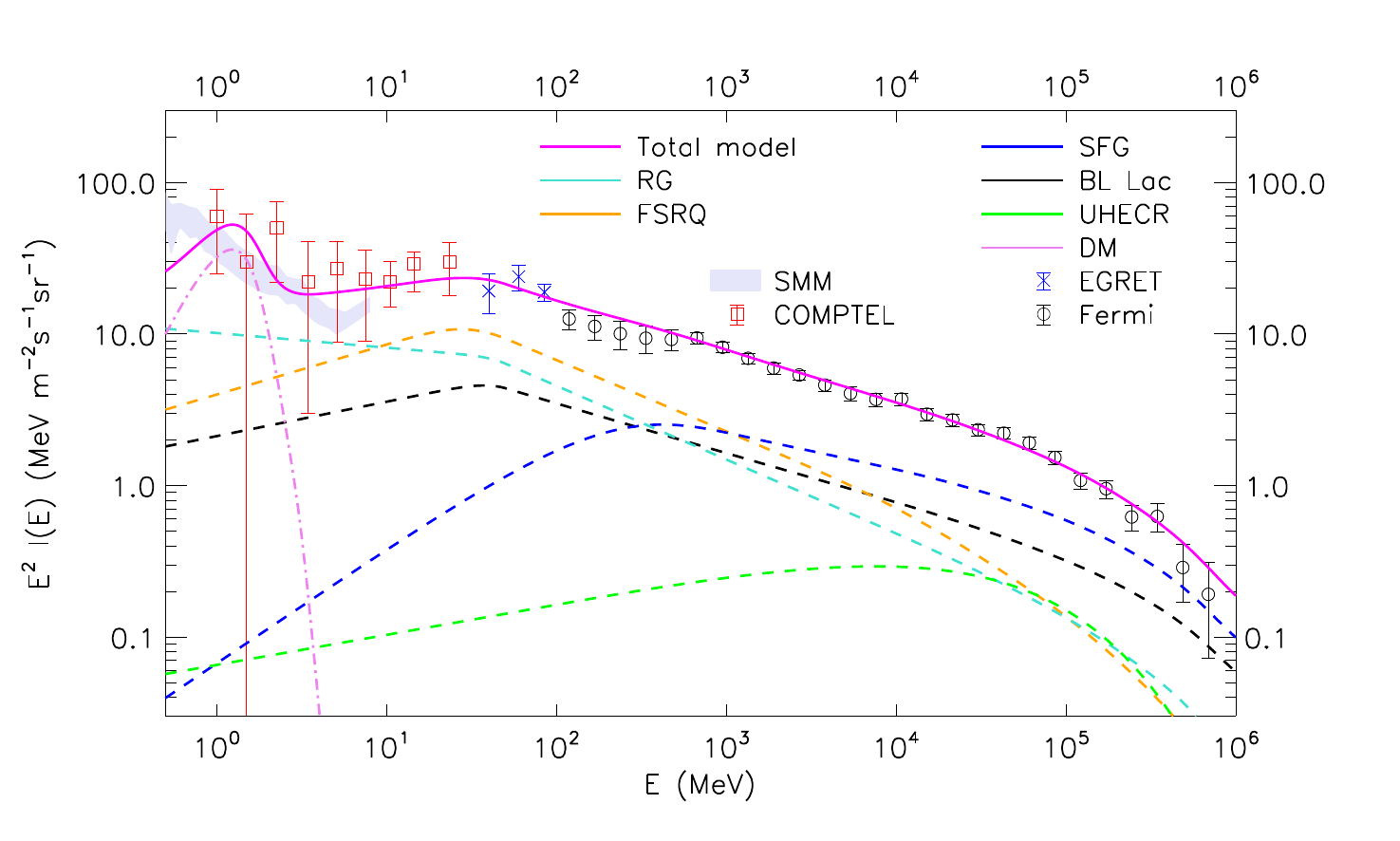}
\end{centering}
\vspace{-0.7cm}
\caption{
The contribution of all components to the EGRB. 
The top panel shows a fit to the EGRB spectrum with a dark matter ``DM'' component from monochromatic PBHs with a mass of $7.4 \times 10^{14}$ g at $z=0$ (dashed-dotted violet line), with an abundance $f_{\textrm{PBH}} = 6.6 \times 10^{-9}$ (best-fit value). In the middle panel, we show a fit to the EGRB spectrum with PBHs of mass of $3.2 \times 10^{15}$ g at $z=0$ , with $f_{\textrm{PBH}} = 7.1 \times 10^{-6}$, and in the bottom panel, we show the fit for PBHs of mass of $4.1 \times 10^{16}$ g at $z=0$, with $f_{\textrm{PBH}} = 6.0 \times 10^{-2}$. In all cases we get a $\chi^{2}$/dof = 0.9-1.0.}
\label{fig:BestFitExamples}
\end{figure}

In Fig.~\ref{fig:BestFitExamples}, we show three examples of a fit to the EGRB data.
While for some choices $m_{\textrm{PBH}}$ of a PBH monochromatic mass-distribution there is no statistical preference for a PBH gamma-ray flux component in certain ranges including these three cases, we have found a small indication for a positive contribution from PBHs. 
The dark matter component ``DM'' in these figures is drawn with a violet dashed-dotted line while the other background astrophysical components are shown with the same color choices as in Fig.~\ref{fig:BackgroundModel}. We show results for PBH masses at $z=0$ of $7.4 \times 10^{14}$ g (top), $3.2 \times 10^{15}$ g (middle), and $4.1 \times 10^{16}$ g (bottom).    
Notice that the PBHs' component has a distinctly different spectrum than the conventional astrophysical components, which allows us to separate it out in the EGRB spectrum and derive tight constraints on it. 

\begin{figure}[h]
\begin{centering}
\hspace{-0.0cm}
\includegraphics[width=3.6in,angle=0]{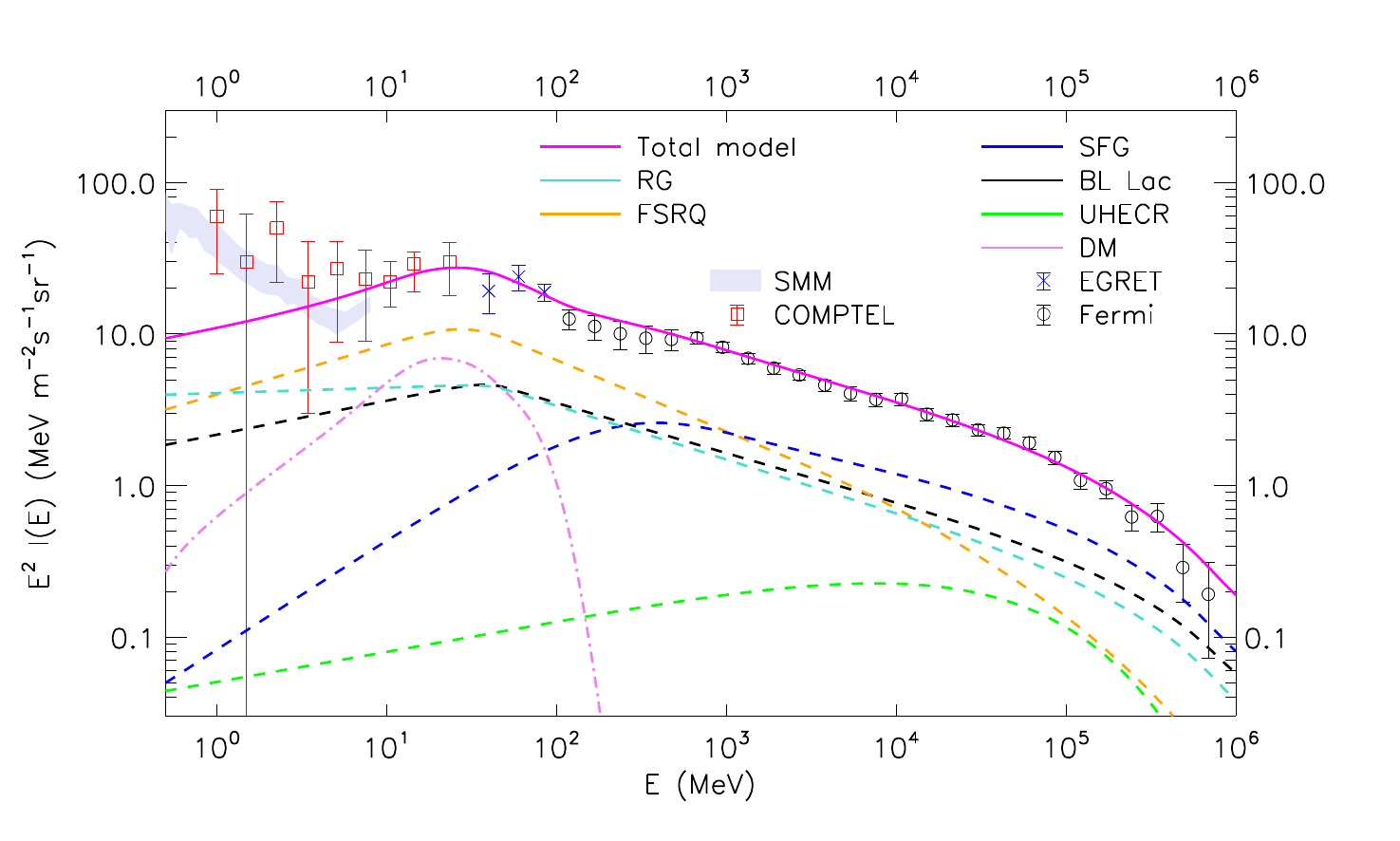}
\end{centering}
\vspace{-0.7cm}
\caption{As in Fig.~\ref{fig:BestFitExamples}, but for  a dark matter ``DM'' component from an extended PBH mass-distribution with peak mass of $m_{\textrm{peak}} = 3.2 \times 10^{15}$ g at formation (dashed-dotted violet line), with an abundance $f_{\textrm{PBH}} = 3.0 \times 10^{-6}$ (best-fit value). We get a $\chi^{2}$/dof = 0.96.}
\label{fig:BestFitExampleForExtendedMassDistr}
\end{figure}

In Fig.~\ref{fig:BestFitExampleForExtendedMassDistr}, we show an example of a fit to the EGRB observations where there is statistical preference for a positive contribution from PBHs that follow an extended mass-distribution. In that example we took the peak of the mass-distribution at the time of formation to be $m_{\textrm{peak}} = 3.2 \times 10^{15}$ g. 
While the gamma-ray spectrum from the extended mass-distribution has a less prominent spectral peak, still such a spectrum is distinctly different from any of the astrophysical background components and thus easily separable in our fits.  

\begin{figure*}[ht]
\begin{centering}
\hspace{-0.2cm}
\includegraphics[width=3.4in,angle=0]{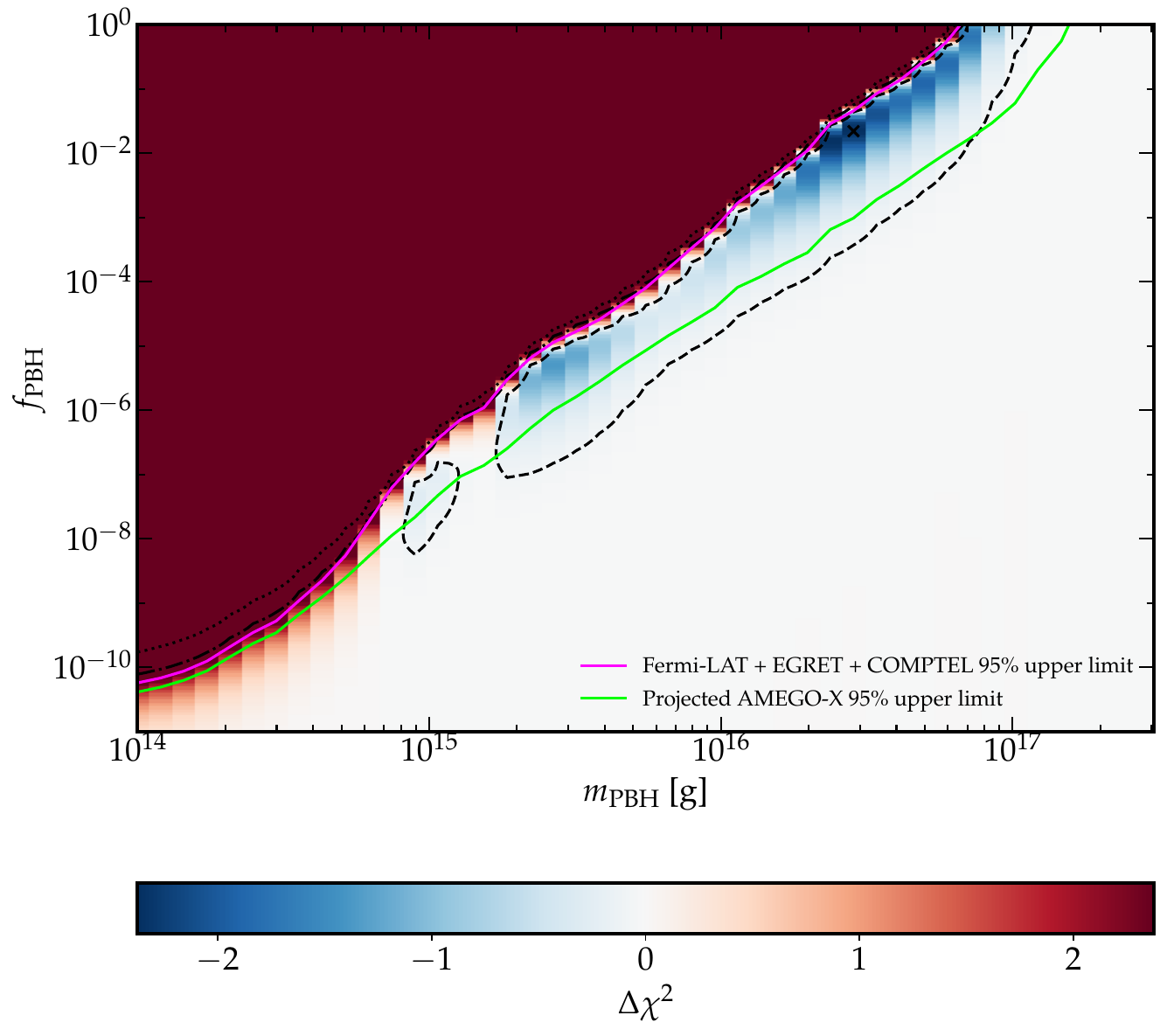}
\hspace{0.5cm}
\includegraphics[width=3.4in,angle=0]{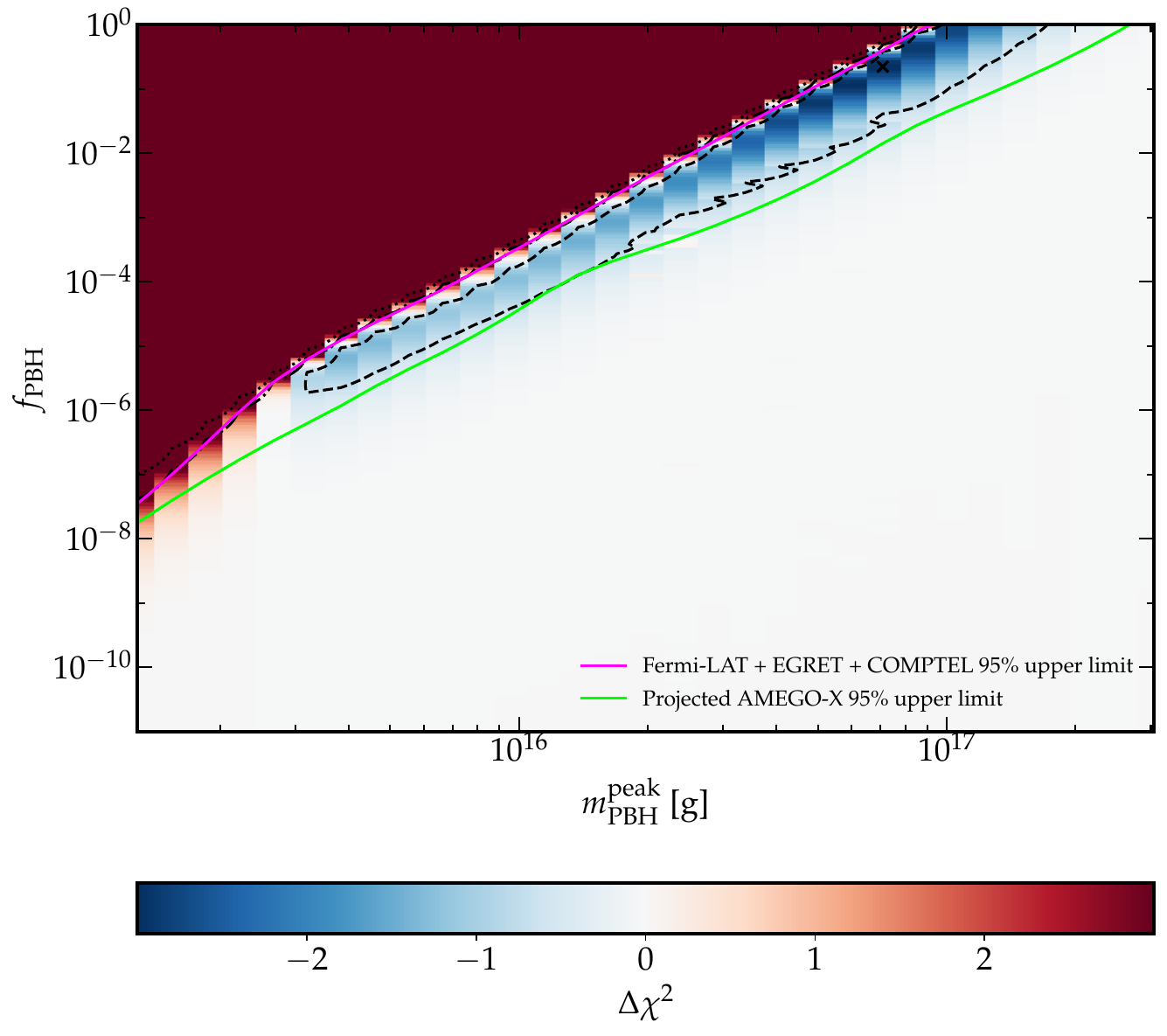}
\end{centering}
\vspace{-0.7cm}
\caption{Using the EGRB observations from the \textit{Fermi}-LAT, \textit{EGRET} and \textit{COMPTEL}, we show our fit results and limits on the abundance of PBHs, $f_{\textrm{PBH}}$. 
\textit{Left}: we take monochromatic PBHs with mass $m_{\textrm{PBH}}$ between $1.0\times 10^{14}$ to $3.1\times 10^{17}$ g.  \textit{Right}: we assume an extended PBH mass-distribution with peak mass $m_{\textrm{peak}}$ in the range of $1.3 \times 10^{15}$ to $3.1 \times 10^{17}$ g.
For every given mass assumption ($m_{\textrm{PBH}}$ or $m_{\textrm{peak}}$) and $f_{\textrm{PBH}}$, we evaluate its $\chi^{2}$ (marginalizing over the background assumptions) and compare this value to the best-fit $\chi^{2}$ derived using only the background components. 
Blue regions ($\Delta \chi^{2} <0$) show combinations of parameters where there is preference for a PBH contribution and red regions ($\Delta \chi^{2} >0$), combinations of parameters where there is a penalty. 
For a value of $f_{\textrm{PBH}} = 1$ (top edge of the $y$-axis) PBHs account for all the dark matter in the Universe. The black ``x'' gives the best-fit PBH parameters. 
For each panel, the black dashed (dotted-dashed and dotted) lines give the region of PBH parameter space that is within $1-\sigma$  ($2-\sigma$ and $3-\sigma$), from the best-fit PBH assumptions (the point ``x'').  
The magenta solid line gives the $95 \%$ upper limit on the $f_{\textrm{PBH}}$. With the current observations we find a slight positive preference for a PBH contribution in the mass range of $m_{\textrm{PBH}} \gsim 2\times 10^{15}$ g and $m_{\textrm{peak}} \gsim 3\times 10^{15}$ g. In the green solid line we also give the projected $95 \%$ upper limits from AMEGO-X assuming no underlying PBH contribution to the EGRB (see text for details).}
\vspace{-0.6cm}
\label{fig:DMlimits}
\end{figure*}

In Fig.~\ref{fig:DMlimits}, we show our limits to the allowed contribution of PBHs to the observed dark matter abundance by fraction of mass $f_{\textrm{PBH}}$. In the left panel of Fig.~\ref{fig:DMlimits}, we show our limits for the case of a monochromatic distribution while on the right panel of the same figure, the limits for an extended mass distribution. 
For any given combination of monochromatic PBH mass $m_{\textrm{PBH}}$ or mass-distribution with peak mass $m_{\textrm{peak}}$, and $f_{\textrm{PBH}}$ abundance, we evaluate the dark matter contribution to the EGRB. We marginalize within the ranges of Table~\ref{tab:FitFreedom}, over the background components' normalizations and low-energy and high-energy spectral indices, to calculate a $\chi^{2}$ for that combination of PBH parameters. 
We compare this $\chi^{2}$ value what we derived assuming no PBHs present. 
Our blue regions are for a $\Delta \chi^{2}<0$ which indicate a preference for a contribution from PBHs to the EGRB combined data. 
Instead,  our red regions for $\Delta \chi^{2}>0$ values show the statistical penalty for adding a PBH component. 

We find only a slight positive preference for a PBH contribution to the EGRB, giving an improvement $\Delta \chi^{2}$ of up to 3, for a PBH contribution with $m_{\textrm{PBH}} \simeq (3-4)\times 10^{16}$ g and an abundance of $f_{\textrm{PBH}} \simeq 6\%$ (or $m_{\textrm{peak}} \simeq 7\times 10^{16}$ g with $f_{\textrm{PBH}} \simeq 20\%$). 
These are depicted by the ``x'' points in the left and right panels of Fig.~\ref{fig:DMlimits}, and are the best-fit PBH parameters for the monochromatic and extended mass-distributions respectively. 
The small preference for a PBH component however is present for wide mass ranges of $m_{\textrm{PBH}} \gsim 2\times 10^{15}$ g and $m_{\textrm{peak}} \gsim 3\times 10^{15}$ g.
We also present with the black dashed, dotted-dashed and dotted lines the region of PBH parameter space that is within 1-$\sigma$, 2-$\sigma$ and 3-$\sigma$ ranges respectively, from the best fit PBH assumptions.  

Finally, the magenta solid line in Fig.~\ref{fig:DMlimits}, gives the $95 \%$ upper limit on the $f_{\textrm{PBH}}$. With the current EGRB observations we cannot place any limits on the abundance of monochromatic PBHs with a mass of $m_{\textrm{PBH}} > 6 \times 10^{16}$ g or PBHs having a mass-distribution with $m_{\textrm{peak}} > 1 \times 10^{17}$ g. 
Such high masses are constrained by the \textit{COMPTEL} data (see Figs.~\ref{fig:BestFitExamples} and~\ref{fig:BestFitExampleForExtendedMassDistr}).
We get a stronger limit on the extended mass distribution parameter $m_{\textrm{peak}}$, as such distributions predict low enough PBH masses that significantly increase the high-energy emitted spectrum; providing enough MeV range gamma rays.  The reader can compare in Fig.~\ref{fig:AltPBHtotalSpectra}, the black solid line to the black long-dashed line that are evaluated for the same value of $m_{\textrm{PBH}}$ and $m_{\textrm{peak}}$ respectively. 

For the extended mass distribution in the left panel of Fig.~\ref{fig:DMlimits}, we stop our derived limits from the EGRB spectrum at $m_{\textrm{peak}} = 1.3 \times 10^{15}$ g, as for lower values of $m_{\textrm{peak}}$ stronger limits can be derived from the lack of close by prominent sources detectable by the \textit{Fermi}-LAT instrument. In particular, assuming a distribution with $m_{\textrm{peak}} = 1 \times 10^{15}$ g, we find that at least one out of every $10^{7}$ PBHs from that distribution, the one that had the lowest mass at formation, approaching its evaporation will emit about $2 \times 10^{25}$ erg/s with a spectrum that has a peak energy of $\sim 1$ GeV. The \textit{Fermi}-LAT instrument has reached a threshold of point source detection of $2 \times 10^{-12}$ erg cm$^{-2}$ s$^{-1}$ \cite{Fermi-LAT:2019yla}. 
That in turn gives us that such a PBH will be detectable if it is within 0.3 pc from the Sun. 
With a local dark matter density of $\simeq 1 \times 10^{-2} \; M_{\odot}/\textrm{pc}^{3}$ \cite{Catena:2009mf, Salucci:2010qr, Bovy:2012tw, Pato:2015dua, deSalas:2020hbh}, for $m_{\textrm{peak}} = 1 \times 10^{15}$ g there are $\sim 2 \times 10^{16}$ PBHs/$\textrm{pc}^{3}$ or $5 \times 10^{14}$ PBHs within 0.3 pc from the Sun. Of them 1 in $10^{7}$ would be detected by \textit{Fermi}-LAT.  
From that we estimate a limit on $f_{\textrm{PBH}} \sim 2 \times 10^{-8}$ for $m_{\textrm{peak}} = 1 \times 10^{15}$ g. 
For an extended mass distribution with $m_{\textrm{peak}} = 5 \times 10^{14}$ ($2.5 \times 10^{14}$) g, the observable radius increases to 100 pc (1 kpc) and the derived limit becomes $f_{\textrm{PBH}} \sim 10^{-13}$ ($5 \times 10^{-16}$). 

\begin{figure}
\begin{centering}
\includegraphics[width=3.35in,angle=0]{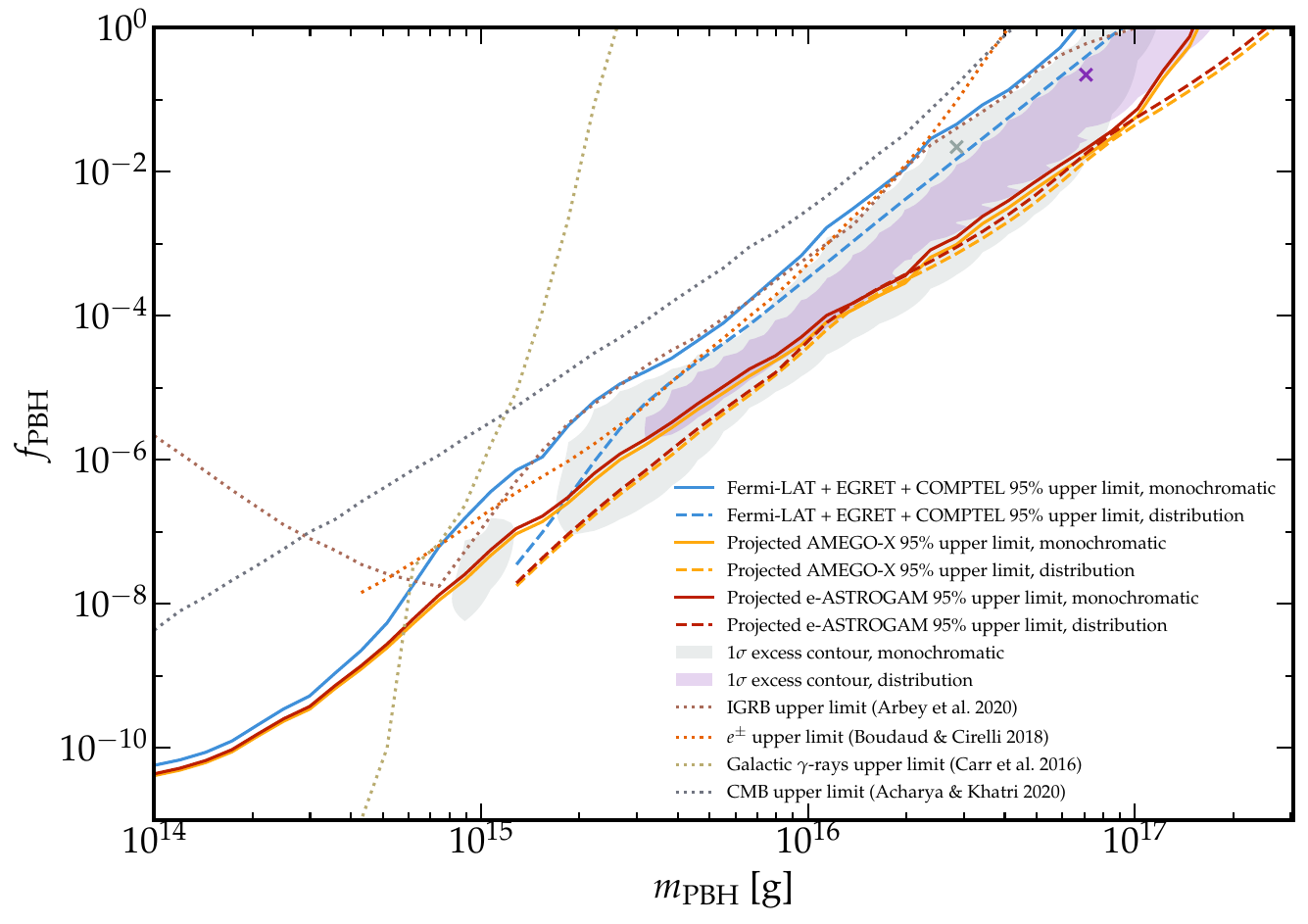}
\end{centering}
\vspace{-0.4cm}
\caption{
Comparison of this work's current and projected EGRB limits on monochromatic PBHs (solid lines), hint for an excess (colored gray region) and other limits in the literature (dotted lines). We present 2-$\sigma$ upper limits from the isotropic gamma-ray background ``IGRB'' \cite{Arbey:2019vqx}, from the cosmic-ray electrons and positrons \cite{Boudaud:2018hqb}, from Galactic gamma-ray observations \cite{Carr:2016hva}, and from the cosmic microwave background anisotropy measurement ``CMB'' \cite{Acharya:2020jbv, Chluba:2020oip}. For ease of comparison in the dashed lines we show the current and projected limits for an extended mass distribution with $m_{\textrm{peak}} = m_{\textrm{PBH}} \geq 1.3 \times 10^{15}$ g. We also show the hint for an excess contribution to the EGRB spectrum from the extended PBH mass distribution (purple region). The ``x'' points give the best-fit parameters for the monochromatic and the extended mass-distributions. }
\label{fig:EGRB_and_other_limits}
\end{figure}

In the literature typically limits are derived under the assumption of a monochromatic PBH mass distribution. For an extended discussion on PBH limits see Ref.~\cite{Carr:2009jm} and the references therein. More recently limits on the PBHs abundance are also discussed for more model-depended extended mass-distributions as e.g. in Ref.~\cite{Carr:2016drx}. 
To compare our limits to the existing limits in the relevant mass range, in Fig.~\ref{fig:EGRB_and_other_limits}, we show our  $95\%$ upper limits on monochromatic PBHs (solid blue line), together with the current limits on monochromatic PBH distributions from other analyses of the isotropic gamma-ray background \cite{Arbey:2019vqx} (brown dotted line),
from the locally measured interstellar medium cosmic-ray electrons and positrons~\cite{Boudaud:2018hqb} (orange dotted line),
from the gamma-ray emission from the Milky Way\cite{Carr:2016hva} (olive dotted line), and from the study of anisotropies in the cosmic microwave background~\cite{Acharya:2020jbv, Chluba:2020oip} (gray dotted line).
Our limits most directly compare to the isotropic gamma-ray background limits from Ref.~\cite{Arbey:2019vqx}, that used similar data. 
They are in fairly good agreement in the mass range of $m_{\textrm{PBH}} : 1\times10^{15}- 1\times 10^{17}$ g; however we derive significantly tighter constraints in the lower mass range as we take a more realistic modeling of the EGRB spectrum accounting for the background astrophysical contributions. Compared to the limits from the study of the Galactic emission~\cite{Carr:2016hva}, we get stronger limits at all masses, comparable only around $10^{15}$ g; with the Galactic emission limits being tighter at $m_{\textrm{PBH}} \simeq 5 \times 10^{14}$ g.  
The limits on $f_{\textrm{PBH}}$ from the locally measured electrons and positrons by \textit{Voyager 1} of Ref.~\cite{Boudaud:2018hqb}, are approximately a factor of 2 stronger in the mass range of $m_{\textrm{PBH}} : 1\times10^{15}- 1\times 10^{16}$ g, but weaker at lower and higher masses. 
Finally, the cosmic microwave background anisotropy limits of Refs.~\cite{Acharya:2020jbv, Chluba:2020oip}, are weaker by a factor of 2-100 across the mass range presented.  
In Fig.~\ref{fig:EGRB_and_other_limits}, by the blue dashed line we also show just for comparison the limits on an extended PBH mass distribution, in which case the $m_{\textrm{PBH}}$ label on the $x$-axis should be understood as $m_{\textrm{peak}}$. 
As we find a small hint for an excess contribution from PBHs in our analysis, we present it for both the monochromatic case (gray region) and the extended mass distribution case (smaller in size purple region). 

For the limits on Figs.~\ref{fig:DMlimits} and~\ref{fig:EGRB_and_other_limits}, we used the EGRB model A of the \textit{Fermi}-LAT measurement. Our limits and best-fit results for the EGRB models B and C of the \textit{Fermi} Collaboration of Ref.~\cite{Fermi-LAT:2014ryh}, are effectively unchanged \footnote{The EGRB spectrum of Ref.~\cite{Fermi-LAT:2014ryh}, was derived using the first 50 months of \textit{Fermi}-LAT observations, when by now we are on the 18th year of data taking. 
A better measurement of the EGRB spectrum is feasible, both by having statistically smaller errors and also possibly reducing the systematic errors associated with modeling and subtracting the contribution of the Milky Way's diffuse emission. }. 
In deriving the presented limits we assumed for the in-flight annihilation conditions similar to the local Milky Way interstellar medium. 
Such an assumption is relevant as the Hawking radiation produced positrons will at a later point annihilate with their surrounding interstellar medium electrons giving gamma rays. 
While at low redshifts the majority of dark matter mass is in large dark matter halos and thus the Milky Way is not an atypical environment where PBHs are located, at higher redshifts most dark matter is in much smaller sized halos (for a discussion regarding the dark matter halo mass-function and its modeling uncertainties see Refs.~\cite{2008ApJ...688..709T, 2011MNRAS.410.1911C, 2013MNRAS.433.1230W, 2017MNRAS.469.4157C, 2021A&A...652A.155S}). 
Thus, we allow also for the environment of PBHs to be one where the baryon density (and thus also electron density with which positrons can annihilate) are present. Even suppressing the in-flight annihilation by a factor of 10 our reported limits on Fig.~\ref{fig:DMlimits} change only to within $O(10\%)$. As the PBH signal is directly proportional to the abundance of dark matter, compared to any dark matter annihilation signal the PBH limits in this work are much less sensitive on the underlying halo mass function.

\section{Using e-ASTROGAM and AMEGO-X to search for signals of primordial black holes}
\label{sec:Projections}

The future AMEGO-X and e-ASTROGAM telescopes will be sensitive in the MeV range and can be used to measure the EGRB; thus also set constraints on the PBH abundance. 

Using the best-fit model EGRB spectrum for the combination of BL Lacs, FSRQs, SFGs, RGs and UHECRs (i.e. with $f_{\textrm{PBH}} = 0$) shown in Fig.~\ref{fig:BackgroundModel}, we evaluated first the expected counts from AMEGO-X and e-ASTROGAM. To do that we used the expected sensitivity curves from  \cite{Fleischhack:2021mhc} and \cite{e-ASTROGAM:2017pxr} respectively. 

Between the two instruments, AMEGO-X is expected to have a better energy resolution, while e-ASTROGAM at certain energies a better angular resolution. We decided to use 34 energy bins for e-ASTROGAM and 54 energy bins for AMEGO-X logarithmically spaced between 0.5 and 1000 MeV. That represents a fractional energy resolution of $25\%$ for e-ASTROGAM  and $15\%$ for AMEGO-X. Both instruments will have energy-dependent energy resolution and not a constant fractional energy resolution. We make a conservative assumption on their capacity. Thus, their limits may be stronger than our prediction. Both e-ASTROGAM and AMEGO-X are going to have a very good sensitivity. Subsequently, the expected counts of photons at the energies of interest will be large enough to lead to fractional systematic errors on the EGRB flux, anywhere between $10^{-4}$ to $10^{-2}$, for both telescopes. However, even the \textit{Fermi}-LAT has a high sensitivity. That did not translate to $<0.5 \times 10^{-2}$ fractional errors on the evaluation of EGRB. In fact, the reported \textit{Fermi}-LAT EGRB spectrum of Ref.~\cite{Fermi-LAT:2014ryh}, at no energy has a fractional error less than $7.5 \times 10^{-2}$.

While alternative cuts on the quality of the \textit{Fermi}-LAT detected events used for an analysis, can affect the \textit{Fermi} flux fractional errors increasing them to $\sim 10^{-2}$, the more important errors are the analysis systematic errors. In particular how from the entire observed gamma-ray flux, the contribution from the foreground diffuse emission from the Milky Way and from the galactic point sources is modeled and removed. 
By the time the AMEGO-X or the e-ASTROGAM telescopes observe the gamma-ray sky some of these analysis systematics are going to improve. However, to be conservative we assume fractional errors on the EGRB expected spectrum, to be $7 \times 10^{-2}$, for either instrument's measurement throughout the 0.5-1000 MeV range. Above $\simeq 1$ GeV the \textit{Fermi}-LAT observation is still going to remain the better one. 

From Fig.~\ref{fig:DMlimits}, we see there are parts of the PBH parameter space that slightly favor a PBH flux component. If a PBH component of mass $m_{\textrm{PBH}} \approx 3\times 10^{16}$ g with an abundance of $f_{\textrm{PBH}} \simeq 5\%$ (or $m_{\textrm{peak}} \gsim 7\times 10^{16}$ g with $f_{\textrm{PBH}} \simeq 20\%$) is present, then
AMEGO-X and/or e-ASTROGAM could detect it. Alternatively, either of these instruments can provide tighter limits on $f_{\textrm{PBH}}$, by a factor of 10 for masses $\gsim 10^{15}$ g and as much as a factor of $10^{2}$ for masses $\gsim 3 \times 10^{16}$ g. This is the result of these higher PBH masses contributing dominantly at energies where now only \textit{COMPTEL} or \textit{EGRET} provide measurements on the EGRB flux (see Figs.~\ref{fig:BestFitExamples} and~\ref{fig:BestFitExampleForExtendedMassDistr}). For masses less than $5 \times 10^{14}$ g the \textit{Fermi}-LAT instrument observes the peak of the PBH emission. However, even for these lower masses a better measurement of the EGRB spectrum at low energies will allow us to more accurately model the contribution of the astrophysical background components. 
Thus, we expect an improved limit on $f_{\textrm{PBH}}$, for these lower masses by a factor of at least 2 (for $m_{\textrm{PBH}} = 10^{14}$ g). 
In Fig.~\ref{fig:EGRB_and_other_limits}, we show the projected $95\%$ upper limits from AMEGO-X (orange lines) and e-ASTROGAM (red lines), together with the results of the previous section. Solid lines are for the monochromatic and dashed lines for the extended mass distributions.  

\section{Summary and Conclusions}
\label{sec:conclusions}

In this paper we revisit the limits that the EGRB spectral measurements from the \textit{Fermi}-LAT, \textit{EGRET} and \textit{COMPTEL} telescopes can place on PBH dark matter, in terms of their contribution to the total dark matter abundance $f_{\textrm{PBH}}$.
We derive limits both for monochromatic PBHs, and for PBHs that follow an extended mass distribution as should be expected from any physical distribution of primordial curvature fluctuations. For such a distribution of PBH masses we follow Ref.~\cite{Biagetti:2021eep}, generating first their seed primordial curvature perturbations. 

Building upon our earlier work in Ref.~\cite{Cholis:2024hmd}, we separately model the contribution to the EGRB from a sequence of astrophysical populations of sources. Those include BL Lacs and FSRQs (see discussion in Section~\ref{subsec:BLLac-FSRQ}), unresolved star-forming and starburst galaxies, as we describe in Section~\ref{subsec:Starforming}, and radio galaxies (discussed in Section~\ref{subsec:Radio}), which as we show also in this work, can have a major contribution to the observed EGRB spectrum.
Furthermore, we find that ultra-high-energy cosmic rays that interact with the intergalactic infrared background are an important component at the higher energy range of the EGRB spectrum (see discussion of Section~\ref{subsec:UHECRs}). 
Our modeling of each component to the EGRB relies on recent developments from detected extragalactic point sources, that account for alternative choices and source-to-source variation in their emitted gamma-ray spectra, on their luminosity distribution properties and in their redshift distribution properties. Moreover, for the modeling of these classes of sources at the MeV energy scale, we rely on observations of specific targets from X-rays, the visible spectrum, the infrared and radio waves. 

We combine all these known astrophysical components and fit them to the EGRB spectrum over the wide energy range of 0.5 MeV to 1 TeV, that the combination of the \textit{COMPTEL}, \textit{EGRET} and \textit{Fermi} telescopes allows us. This is shown in Fig.~\ref{fig:BackgroundModel}. We find that each one of these emission sources at some range of energies within these more than six orders of magnitude in gamma-ray energy, can be the dominant or a nearly dominant component.

In our fits we allow for freedom in the normalizations of each of the five astrophysical components described; while we also allow for a range of values for each population's of sources (BL Lacs, FSRQs, star-forming $\&$ starburst galaxies and the radio galaxies) combined low-energy and high-energy spectral indices. 
These degrees of freedom are motivated by the need to properly account for the existing modeling uncertainties of those populations and also allow us to produce conservative limits on PBH dark matter (see discussion  about the fitting procedure in Sections~\ref{subsec:Combination} and~\ref{subsec:analysis}).
Within a reasonable modeling degree of freedom in combining the known astrophysical components, we find good fits of $\chi^{2}$/dof $\simeq 1$ to the EGRB spectrum. 

In modeling the contribution from PBHs to the EGRB observations (discussed in Section~\ref{subsec:DM}), we find that the dominant contribution is indeed of extragalactic origin, i.e. the emission of gamma rays from distant galaxies/halos, while gamma rays produced from PBHs in the Milky Way halo at high latitudes are a small component. We study monochromatic PBHs with mass $m_{\textrm{PBH}}$ at z=0 between $1.0 \times 10^{14}$ and up to $3.1 \times 10^{17}$ g. Using a  public package that we recently developed for that purpose, \texttt{GammaPBHPlotter} of Ref.~\cite{Carlini:2025bki}, we account for i) the direct Hawking radiation, ii) gamma rays from the hadronization and decay of unstable particles, iii) final state radiation and iv) gamma rays from subsequent pair annihilations. The latter two components have been neglected in the literature (see though~\cite{Keith:2022sow}), but can be of importance in setting up accurate limits on PBHs as they significantly increase their low-energy spectrum. In Fig.~\ref{fig:PBHcomponents}, we show how those four components contribute to the total PBH gamma-ray spectrum of a given mass black hole. For each of our mass choices, we account for the mass evolution of the PBHs as their masses gradually decrease, with their emitting power increasing and their gamma-ray spectra shifting to higher energies. We evaluate the entire emission from PBHs from a redshift of $z=10$ to $z=0$ (examples of resulting spectra are given in Fig.~\ref{fig:AltPBHtotalSpectra}). We repeat the same procedure in studying an extended mass distribution of PBHs with a peak of their relevant probability density function at formation $m_{\textrm{peak}}$ between $1.3 \times 10^{15}$ up to $3.1 \times 10^{17}$ g. 

Including a PBH emission component to the EGRB can improve the fit to the spectral data by a small amount in the mass ranges of $m_{\textrm{PBH}} \gsim 2\times 10^{15}$ g and $m_{\textrm{peak}} \gsim 3\times 10^{15}$ g. The significance for a PBH contribution is up to $\Delta \chi^{2} \lesssim 3$, with our overall best fit choices found for $m_{\textrm{PBH}} \simeq 4\times 10^{16}$ g, with $f_{\textrm{PBH}} \simeq 6\%$ and for $m_{\textrm{peak}} \gsim 7\times 10^{16}$ g with $f_{\textrm{PBH}} \simeq 20\%$ (see also Figs.~\ref{fig:BestFitExamples},~\ref{fig:BestFitExampleForExtendedMassDistr} and~\ref{fig:DMlimits}). While these results give only small hint for a PBH contribution, the future AMEGO-X and e-ASTROGAM MeV-scale telescopes will be able to either detect it or exclude it as their observations will allow us to better model the conventional astrophysical populations of sources emitting in these energies and at energies currently observed by the \textit{Fermi}-LAT.     

The current EGRB spectral measurements allow us to place upper limits on the PBH abundance for monochromatic PBHs with $m_{\textrm{PBH}} < 6 \times 10^{16}$ g, and for PBHs having a mass-distribution with $m_{\textrm{peak}} < 1 \times 10^{17}$ (shown in Fig.~\ref{fig:DMlimits}). Our $95\%$ upper limits can be as stringent as $f_{\textrm{PBH}} \simeq 10^{-10}$ ($f_{\textrm{PBH}} \simeq 10^{-8}$) for $m_{\textrm{PBH}} \simeq 1 \times 10^{14}$ ($m_{\textrm{peak}} \simeq 1.3 \times 10^{15}$) g. These are among the tightest limits for PBHs in the mass range of $10^{14}$-$10^{17}$ g. We compare our results to alternative probes of PBHs in the same mass range in Fig.~\ref{fig:EGRB_and_other_limits}.

Our limits are robust to alternative estimations of the EGRB spectrum by the \textit{Fermi} Collaboration of Ref.~\cite{Fermi-LAT:2014ryh}. Furthermore, these limits are also sensitive only to within a few $\%$ in their values on alternative choices on the medium surrounding the PBHs. 

Finally, we note that even with conservative assumptions, the future observations from AMEGO-X and/or e-ASTROGAM  can improve those limits across the entire mass range studied and for the highest masses up to a factor of $\simeq 100$.  

We make publicly available our dark matter and astrophysical background simulation spectral files at \texttt{https://zenodo.org/records/20563575}. 
  
\acknowledgements
We acknowledge the use of \path{Python} \cite{10.5555/1593511} modules, \path{numpy} \cite{harris2020array},
\path{SciPy} \cite{2020SciPy-NMeth}, \path{matplotlib} \cite{Hunter:2007}, \path{Jupyter} \cite{Kluyver2016jupyter}, and \path{iminuit} \cite{iminuit,James:1975dr}.
IC acknowledges that this material is based upon work supported by the U.S. Department of Energy, Office of Science, 
Office of High Energy Physics, under Award No. DE-SC0022352.

%\begin{appendix}

%\section{The impact of background modeling freedom on the dark matter limits}
%\label{app:background_scrict_fit}

%\end{appendix}  

\bibliography{EGRB_PBH}

\end{document}